\documentclass[draftclsnofoot,onecolumn,12pt]{IEEEtran}
\usepackage{algorithm,amsbsy,amsmath,amssymb,epsfig,bbm,mathrsfs,multirow,amsthm}
\usepackage{array,multirow,graphicx}
\usepackage{graphicx}
\usepackage{graphics}
\usepackage{epstopdf}
\usepackage{color}
\usepackage{bbm}
\usepackage{comment}
\usepackage{cite}
\usepackage{algpseudocode}
\usepackage{mathtools,xparse}
\usepackage{subcaption}
\usepackage{mathtools}
\definecolor{mygray}{gray}{0.6}
\definecolor{myblue}{rgb}{0.8,0.85,1} 
\usepackage{array}
\newcolumntype{L}[1]{>{\raggedright\let\newline\\\arraybackslash\hspace{0pt}}m{#1}}
\newcolumntype{C}[1]{>{\centering\let\newline\\\arraybackslash\hspace{0pt}}m{#1}}
\newcolumntype{R}[1]{>{\raggedleft\let\newline\\\arraybackslash\hspace{0pt}}m{#1}}

\usepackage{tabularx, boldline}
\usepackage{cellspace}

\usepackage{grffile}

\makeatletter
\def\ScaleIfNeeded{%
\ifdim\Gin@nat@width>\linewidth \linewidth \else \Gin@nat@width \fi
} \makeatother

\newcommand{\defequal}{\mbox{$\stackrel{\triangle}{=}$}}

\newcommand{\ds}{\displaystyle}

\newcommand{\clN}{{\cal N}}

\newcommand{\clK}{{\cal K}}

\newcommand{\clO}{{\cal O}}
\newcommand{\clU}{{\cal U}}

\newcommand{\bh}{\mathbf{h}}
\newcommand{\bg}{\mathbf{g}}

\newcommand{\bG}{\mathbf{G}}

\newcommand{\bw}{\mathbf{w}}

\newcommand{\bt}{\mathbf{t}}

\newcommand{\bp}{\mathbf{p}}

\newcommand{\bv}{\mathbf{v}}

\newcommand{\noise}{\sigma^2}

\newcommand{\ti}{t_i}
\newcommand{\te}{t_e}
\newcommand{\td}{t_d}
\newcommand{\tni}{\tau_i}
\newcommand{\tne}{\tau_e}
\newcommand{\tnd}{\tau_d}
\newcommand{\tnik}{\tau_i^{(\kappa)}}
\newcommand{\tnek}{\tau_e^{(\kappa)}}
\newcommand{\tndk}{\tau_d^{(\kappa)}}

\newcommand{\la}{\langle}
\newcommand{\ra}{\rangle}

\newcommand{\vk}{v^{(\kappa)}}
\newcommand{\wk}{w^{(\kappa)}}
\newcommand{\pk}{p^{(\kappa)}}
\newcommand{\tauk}{\tau^{(\kappa)}}
\newcommand{\vkk}{v^{(\kappa+1)}}
\newcommand{\wkk}{w^{(\kappa+1)}}
\newcommand{\pkk}{p^{(\kappa+1)}}
\newcommand{\taukk}{\tau^{(\kappa+1)}}
\newcommand{\brc}{\boldsymbol{\Theta}}
\newcommand{\brcv}{\boldsymbol{\theta}}
\newcommand{\rcvk}{\theta^{(\kappa)}}
\newcommand{\rcvkk}{\theta^{(\kappa+1)}}

\newcommand{\btau}{\boldsymbol{\tau}}
\begin{document}
	
\title{Intelligence Reflecting Surface-Aided Integrated Data and Energy Networking Coexisting D2D Communications}

\author{
	{
Nguyen Thi Thanh Van, Huy T. Nguyen, 
Nguyen Cong Luong, Ngo Manh Tien,
Dusit Niyato,~\IEEEmembership{Fellow,~IEEE}
and
Dong In Kim,~\IEEEmembership{Fellow,~IEEE}
	}
}

\maketitle

\thispagestyle{empty}

\begin{abstract}
In this paper, we consider an integrated data and energy network and D2D communication coexistence (DED2D) system. The DED2D system allows a base station (BS) to transfer data to information-demanded users (IUs) and energy to energy-demanded users (EUs), i.e., using a time-fraction-based information and energy transfer (TFIET) scheme. Furthermore, the DED2D system enables D2D communications to share spectrum with the BS. Therefore, the DED2D system addresses the growth of energy and spectrum demands of the next generation networks. However, the interference caused by the D2D communications and propagation loss of wireless links can significantly degrade the data throughput of IUs. To deal with the issues, we propose to deploy an intelligent reflecting surface (IRS) in the DED2D system. Then, we formulate an optimization problem that aims to optimize the information beamformer for the IUs, energy beamformer for EUs, time fractions of the TFIET, transmit power of D2D transmitters, and reflection coefficients of the IRS to maximize IUs' worse throughput while satisfying the harvested energy requirement of EUs and D2D rate threshold. The max-min throughput optimization problem is computationally intractable, and we develop an alternating descent algorithm to resolve it with low computational complexity. The simulation results demonstrate the effectiveness of the proposed algorithm.
\end{abstract}

\begin{IEEEkeywords}
Intelligence reflecting surface, energy harvesting, data transfer, D2D communication, noncovex optimization problem.
\end{IEEEkeywords}

\section{Introduction}\label{section:2}

The growth of the next generation networks such as IoT systems raises challenges of energy and spectrum demands~\cite{hu2018integrated}. To address the energy challenge, integrated data and energy networks have been recently proposed in which a base station (BS) can perform information transfer to information-demanded users (IUs) and energy transfer to energy-demanded users (EUs). Meanwhile, to address the scarcity of spectrum resource and reduce the energy consumption, Device-to-Device (D2D) communication is still known as an effective solution that allows D2D transmitters to share spectrum with cellular networks~\cite{xu2012interference} for a short communication range. As a result, the integrated data and energy network and D2D communication coexistence (DED2D) is a promising combination that is able to address the challenges of the demand growth of energy and spectrum resources in the IoT system as well as in the next generation networks. 

However, the DED2D system raises other issues. The first issue comes from the coexistence of the D2D communications and the integrated data and energy network. In particular, the D2D transmitters share the frequency band with the data and energy network, and thus the D2D communication causes the co-channel interference to the IUs and significantly degrades the data throughput of IUs~\cite{xu2012interference}. The second issue is that the wireless channel impairments, e.g., the path loss and multi-path fading, significantly degrade the data throughput of IUs. The third issue is that EUs are typically IoT devices that require substantially harvested energy, and the harvested energy substantially decreases when the distance from the BS to EUs is long. To improve the data throughput, some technologies, e.g., massive
multiple-input multiple-output (MIMO), have recently been proposed. However, they require high energy consumption and hardware cost~\cite{IEEE_IRS_HDTuan}.


Intelligent Reflecting Surface (IRS) has recently been proposed as an emerging technology for the development of the next-generation wireless networks~\cite{wu2019towards}, \cite{di2019smart}. An IRS consists of low-cost passive elements that can reflect incident signals by intelligently adjusting their phase-shifts corresponding to wireless channels. The signals reflected by the IRS can be added constructively with non-reflected signals, i.e., Line-of-Sight (LoS) signals, at receivers to boost the received signal power, or can be destructively added to suppress the co-channel interference~\cite{cai2020two}, \cite{zhou2020intelligent}. As such, IRS is a promising solution to deal with the aforementioned issues, i.e., the co-channel interference, propagation loss, and obstacles, of the DED2D system. It is also well presented in~\cite{wu2019weighted} that deploying IRS around EUs can significantly improve the total harvesting energy compared with the case without using IRS.   

In this paper, we consider an integrated data and energy network and D2D communication coexistence (DED2D) system with assistance of IRS. In the IRS-aided DED2D system, a BS transfers information to multiple IUs and energy to multiple EUs through the time-fraction-based information and energy transfer (TFIET) scheme~\cite{yu2020improper}. The D2D transmitters can use the same frequency band as that of the integrated data and energy network for their data transmissions. An IRS is deployed in the system with the aim of improving the data throughput of the IUs and D2D communications as well as well managing the interference caused from the D2D communications to the IUs. We investigate an optimization problem that jointly optimizes the i) information beamformer for IUs, ii) energy beamformer for EUs, iii) time fractions of the TFIET scheme, iv) transmit power of the D2D transmitters, and v) reflection coefficients of IRS. The objective is to maximize the minimum throughput among the IUs while satisfying the energy requirement of the EUs and the D2D communication rate threshold. The optimization problem is computationally intractable due to the fact that both the objective and constraints are complex functions of beamformers and IRS reflecting coefficients. Especially, the reflecting coefficients of the IRS are constrained by the nonconvex unit-modulus constraint.

To the best of our knowledge, there is no work investigating the max-min throughput optimization problem in the IRS-aided DED2D system, which is a combination of emerging technologies, i.e., IRS, integrated data and energy network, and D2D communication. In particular, there is a number of works that propose to use IRS to improve the data throughput in wireless networks. The readers are referred to a comprehensive survey of such works in \cite{gong2020toward}. Recently, there have some works related to IRS-aided D2D communications~\cite{cao2021sum}, \cite{cai2020two},~\cite{mao2021intelligent}, \cite{nguyen2021deep}. Specifically, the authors in~\cite{cao2021sum} aim to maximize the sum-rate of a D2D pair and an IU in the IRS-aided D2D communication system by jointly optimizing the transmit power at the D2D transmitter and phase shifts of IRS. To reduce the required channel training and feedback overhead, the authors in \cite{cai2020two} propose a two-timescale optimization that aims to optimize the transmit beamforming at the BS, transmit power at the D2D transmitter, and IRS phase shifts, subject to the outage probability of the IU. Different from ~\cite{cao2021sum} and \cite{cai2020two}, the authors in~\cite{mao2021intelligent} propose to use the IRS to reduce the offloading latency for the D2D transmitter. As an extension of the aforementioned works, the authors in \cite{nguyen2021deep} consider the IRS-aided D2D communications with multiple D2D pairs, and a  deep reinforcement learning algorithm is adopted to optimize the transmit power of D2D transmitter and phase shifts of the IRS under the dynamics of wireless channels. To improve both the data throughput and harvested energy for the devices, recent works such as \cite{zargari2021max}, \cite{tang2020joint}, \cite{pan2020intelligent}, \cite{wu2019weighted},  \cite{tuan2019non},~\cite{IEEE_IRS_HDTuan}, and~\cite{yu2020improper} propose to deploy IRS in the integrated data and energy networks. In particular, the authors in \cite{zargari2021max} propose to use an IRS for a simultaneous wireless information and power transfer (SWIPT) system. The SWIPT system allows each receiver, i.e., user, to split the received signal into two components, one for energy-harvesting (EH) and one for information decoding (ID) by using the power splitting protocol. Then, the authors jointly design the energy and information beamforming at the BS, phase shifts of the IRS, and power
splitting ratio at all users to maximize the minimum energy efficiency of users subject to the minimum rate and minimum harvested energy. In fact, it is hard to optimize the conflicting targets of information and energy beamforming at the same time. Thus, the users can be divided into IUs and EUs, and then the authors of \cite{tang2020joint} jointly optimize the information beamformer for IUs, the energy beamformer for the EUs, and phase shifts of the IRS. In \cite{pan2020intelligent}, the authors jointly optimize the information beamformer for the IUs and the phase shift of IRS to maximize the weighted sum rate of IUs while simultaneously satisfying the energy harvesting requirement of the EUs. The system model and the optimization problem in \cite{pan2020intelligent} can be found in~\cite{wu2019weighted}, but the work in~\cite{wu2019weighted} aims to maximize the weighted sum-power received at EUs, subject to the SINR constraints for the IUs. The simulation results in~\cite{wu2019weighted} show that the sum-power obtained by EUs significantly improves than that in the case without IRS. However, these works do not consider fairness in their optimizations.

As mentioned earlier, the optimization problems for the beamformers and the reflecting coefficients of IRS are generally very challenging. Recent works, e.g., in \cite{tuan2019non},~\cite{IEEE_IRS_HDTuan}, and~\cite{yu2020improper}, have shown the effectiveness of alternating optimization algorithms that can solve the optimization problems with low computational complexity and converge at least to a locally optimal solution. Therefore, in this work, we develop an alternating optimization algorithm to solve the max-min throughput problem in the IRS-aided DED2D system. The main contributions of the paper include the followings:

\begin{itemize}
\item This is the first work that considers the use of IRS for the integrated data and energy network and D2D communication coexistence system. The integrated data and energy network allows the BS to perform the information transfer to multiple IUs and the energy transfer to multiple EUs by using the TFIET scheme, while the D2D communications aim to improve the spectrum efficiency. The IRS is deployed for supporting the data transfer of the EUs and data transmission of the D2D communications as well as managing the interference caused by the D2D communications to the IUs.
\item We formulate an optimization problem for the IRS-aided DED2D system that jointly optimizes the information beamformer for the IUs, energy beamformer for the EUs, time fractions of the TFIET scheme, transmit power of D2D transmitters, and reflection coefficients of IRS, so as to maximize the minimum throughput among the IUs, subject to the energy requirement of the EUs and the data rate threshold of the D2D communications. 
\item Despite the objective to achieve fairness, the throughput problem is computationally intractable. Thus, we develop an alternating descent algorithm that iteratively solves the problem with efficient computation. This algorithm is proposed for the scenario in which there is no orthogonal time allocation (N-OTA) between the D2D communications and the data and energy transfers, and thus it is namely N-OTA algorithm.
\item We further consider a scenario, namely OTA, in which the BS performs the OTA among the D2D communications and the data and energy transfers. In this case, there are no interference caused by the D2D communications to the integrated data and energy network, but the time for the data and energy transfers reduces. Then, we formulate the max-min throughput for the IRS-aided DED2D system in which the time allocation to the D2D communications is included. To solve this computationally intractable problem, we again use the alternating descent algorithm. 
\item We provide simulation results to show the improvement of the proposed algorithms, i.e., the OTA and N-OTA algorithms, compared with the baseline algorithms in which the IRS phase shifts are random. The results further reveal some interesting findings. For example, the efficiency of the OTA-algorithm over the N-OTA algorithm can change depending on the D2D rate threshold and the IRS size.
\end{itemize}

The rest of the paper is organized as follows. In Section~\ref{section:3}, we describe the IRS-aided DED2D system and formulate the optimization problem for the system. In Section~\ref{path-follo-prob1}, we present the alternating descent algorithm proposed to solve the problem. In Section~\ref{sec:ota}, we formulate the optimization problem in the IRS-aided DED2D system with OTA scenario and present the alternating descent algorithm to solve it. The simulation
results and discussions are presented in Section~\ref{perform_eval}, and the conclusions are given in Section~\ref{sec:conc}.
\section{System Model and Problem Formulation}
\label{section:3}
\begin{figure}[b!]
		\centerline{\includegraphics[width=0.75\textwidth]{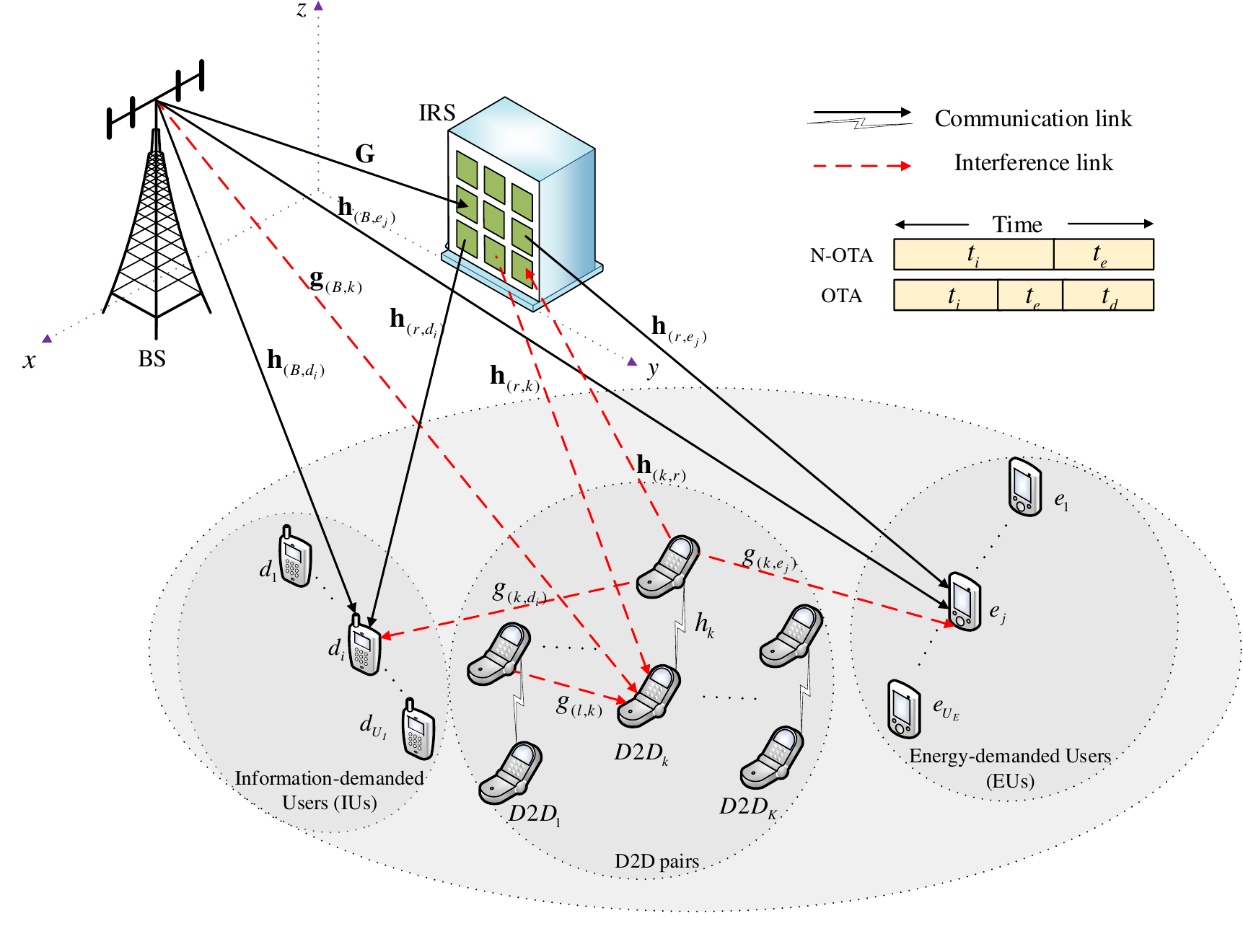}}
		\caption{The IRS-aided integrated data and energy network.}
	\label{fig:model}
	\vspace{-1cm}
\end{figure}

\subsection{System Model}
\label{sec:sys}
We consider an IRS-aided DED2D network as shown in Fig.~\ref{fig:model}. The BS is equipped with $M$ antennas to serve a set $\clU_I \defequal \{(d_i), i=1,\dots,U_I\}$ of IUs and a set $\clU_E \defequal \{(e_j), j = 1,\dots,U_E\}$ of EUs. In this work, we consider the downlink scenario in which IUs receive the data transmitted from BS, and EUs harvest the energy from the signals transmitted from BS. Both IUs and EUs are equipped with single-antenna and operate in half-duplex mode. There is a set $\clK=\{1,\dots,K\}$ of D2D pairs that share the same spectrum with the primary network. The IRS is equipped with a set $\clN=\{1,\dots,N\}$ of reflecting elements. Note that the network can be extended with multiple IRSs, and the problem remains significantly unchanged. Let $\brc \in \mathbb{C}^{N \times N}$ denote the diagonal matrix of reflection coefficients. We have $\brc \triangleq \text{diag} (\theta_1, \theta_2,\ldots, \theta_N)$, where $\theta_n$ is the reflection coefficient of element $n$ of IRS, which satisfies the following uni-modulus condition: 
\begin{equation} \label{umc}
|\theta_n| = 1, \forall n\in\clN.
\end{equation}

Let $\bG \in\mathbb{C}^{N \times M}$ denote the channel from BS to IRS. We denote $\bh_{(k,r)}\in\mathbb{C}^{N}$ as the reflect interference channel from D2D transmitter of pair $k$ to IRS. Also, we denote $\bh_{(B,d_i)}\in\mathbb{C}^{M}$, $\bh_{(B,e_j)}\in\mathbb{C}^{M}$, and $\bg_{(B,k)}\in\mathbb{C}^{M}$ as the direct channels from BS to IU $d_i \in \clU_I$, EU $e_j \in \clU_E$, and the interference channel from BS to D2D receiver of pair $k \in \clK$, respectively. Furthermore, we denote $\bh_{(r,d_i)}\in\mathbb{C}^{N}$, $\bh_{(r,e_j)}\in\mathbb{C}^{N}$, and $\bh_{(r,k)}\in\mathbb{C}^{N}$ as the channels from IRS to IU $d_i \in \clU_I$, EU $e_j \in \clU_E$, and D2D receiver of pair $k \in \clK$, respectively. 

For the D2D communications, we denote $h_{k}\in\mathbb{C}$ and $g_{(l,k)}\in\mathbb{C}$ as the communication channel of D2D pair $k$ and interference channel from D2D transmitter of pair $l$ to D2D receiver of pair $k$, respectively. Also, let $g_{(k,d_i)}\in\mathbb{C}$ and $g_{(k,e_j)}\in\mathbb{C}$ denote the interference channels from D2D transmitter of pair $k$ to IU $d_i \in \clU_I$ and EU $e_j \in \clU_E$, respectively.

To perform the data transmission and energy transfer of BS, we adopt the time-switching protocol ~\cite{nasir2016beamforming}. Let $\ti$ and $\te$ be the factions of time used for the data transmission and energy transfer of the BS, respectively, and we have $0 \leq \ti+\te \leq 1$. Note that the D2D transmitters communicate with their receivers during $\ti$ and $\te$. Then, the signals at IUs, EUs, and D2D receivers can be determined as follows.

\subsubsection{Received signals at IUs}
During the data transfer period of $\ti$, the signal received at IU $d_i \in \clU_I$ consists of i) the signal transmitted directly from BS, ii) the signal transmitted from BS and reflected by IRS, iii) the signal transmitted directly from D2D transmitter of pair $k$, and iv) the signal transmitted from the D2D transmitter and reflected by IRS. Let $y_{d_i}$ denote the signal received at  IU $d_i$, then $y_{d_i}$ is determined by
\begin{align}
y_{d_i} = \ds\sum_{l=1}^{U_I} \bh_{(B,d_i)}(\brcv) \bw_{d_l} s_{d_l} + \ds\sum_{k=1}^{K} g_{(k,d_i)}(\brcv) \sqrt{p_k}  s_k  + n_{d_i}, 
\end{align}
where $\bh_{(B,d_i)}(\brcv) = \bh_{(B,d_i)}^H + \bh_{(r,d_i)}^H \brc \bG$, $\bh_{(B,d_i)}(\brcv) \in \mathbb{C}^{1 \times M}$, $g_{(k,d_i)}(\brcv) = g_{(k,d_i)} + \bh_{(r,d_i)}^H \brc \bh_{(k,r)}$, $\bw_{d_l} \in \mathbb{C}^{M}$ and $s_{d_l} \in \mathbb{C}$ are the information beamformer and symbol intended to IU $d_l$, respectively, $p_k$ and $s_k$ are the transmit power of D2D transmitter of pair $k$ and the symbol intended to D2D receiver of pair pair $k$, respectively, and $n_{d_i} \sim \mathcal{CN}(0,1)$ is the zero-mean 
additive white Gaussian noise (AWGN) with unit variance at IU $d_i$.
\subsubsection{Received signals at EUs}
During the energy transfer period of $\te$, the signal received at EU $e_j \in \clU_E$ consists of i) the signal transmitted directly from BS, ii) the signal transmitted from BS and reflected by IRS, iii) the signal transmitted directly from the D2D transmitter of pair $k$, and iv) the signal transmitted from the D2D transmitter reflected by IRS. Let $y_{e_j}$ be the signal received at EU $e_j$. Then, $y_{e_j}$ is determined by
\begin{align}
y_{e_j} = \ds\sum_{l=1}^{U_E} \bh_{(B,e_j)}(\brcv) \bv_{e_l} s_{e_l} + \ds\sum_{k=1}^{K} g_{(k,e_j)}(\brcv) \sqrt{p_k}  s_k^d  + n_{e_j}, 
\end{align}
where $\bh_{(B,e_j)}(\brcv) = \bh_{(B,e_j)}^H + \bh_{(r,e_j)}^H \brc \bG$, $\bh_{(B,e_j)}(\brcv) \in \mathbb{C}^{1 \times M}$, $g_{(k,e_j)}(\brcv) = g_{(k,e_j)} + \bh_{(r,e_j)}^H \brc \bh_{(k,r)}$, $\bv_{e_l} \in \mathbb{C}^{M}$ and $s_{e_l} \in \mathbb{C}$ are the energy beamformer and energy symbol intended to EU $e_l$, respectively, and $n_{e_j} \sim \mathcal{CN}(0,1)$ is the additive white Gaussian noise at EU $e_j$. Here, the energy symbol does not carry any information~\cite{lee2015optimal}.

\subsubsection{Signals at D2D receivers}

As mentioned earlier, the D2D pairs work during both $\ti$ and $\te$. In particular, during $\ti$, the signal at D2D receiver of pair $k \in \clK$ consists of i) the signal transmitted directly from the D2D transmitter of pair $k$, ii) the signal transmitted from the D2D transmitter and reflected by the IRS, iii) the signals transmitted from the D2D transmitters of pairs $l, \forall l \in \clK \setminus \{k\}$, and iv) the signals corresponding to the IUs transmitted from the BS and reflected by the IRS.

The signals at the D2D receivers during $\te$ are formulated similar to those during $\ti$. However, during $\te$, the BS transmits energy symbols rather than the data. Therefore, the signals at the D2D receivers during $\te$ should include the signals corresponding to the EUs transmitted from the BS and reflected by the IRS. Let $y_{\ti,k}$ and $y_{\te,k}$ denote the signal at D2D receiver of pair $k$. Then, $y_{\ti,k}$ and $y_{\te,k}$ can be expressed as follows:
\begin{align}
y_{\ti,k} &= h_{k}(\brcv) \sqrt{p_k} s_k + \ds\sum_{l \in \clK \setminus \{k\}}g_{l,k}(\brcv) \sqrt{p_l} s_l  + \ds\sum_{i=1}^{U_I} \bg_{(B,k)}(\brcv) \bw_{d_i} s_{d_i} + n_{k}, \nonumber \\
y_{\te,k} &= h_{k}(\brcv) \sqrt{p_k} s_k + \ds\sum_{l \in \clK \setminus \{k\}}g_{l,k}(\brcv) \sqrt{p_l} s_l  +  \ds\sum_{j=1}^{U_E} \bg_{(B,k)}(\brcv) \bv_{e_j} s_{e_j} + n_{k},
\end{align}
where $h_{k}(\brcv) = h_{k} + \bh_{(r,k)}^H \brc \bh_{(k,r)}$, $g_{(l,k)}(\brcv) = g_{(l,k)} + \bh_{(r,k)}^H \brc \bh_{(l,r)}$, $\bg_{(B,k)}(\brcv) = \bg_{(B,k)}^H + \bh_{(r,k)}^H \brc \bG$, and $n_{k} \sim \mathcal{CN}(0,1)$ is an additive white Gaussian noise at D2D receiver of pair $k$.

To simplify the presentation, we define the following vectors: $\bw\triangleq \{\bw_{d_i}, d_i \in \clU_I \}, \bv \triangleq \{\bv_{e_j}, e_j \in \clU_E\}, \brcv \triangleq (\theta_1,\dots,\theta_N), \bt \triangleq \{\ti, \te\}$, and $\bp\triangleq \{p_{k}, k \in \clK\}$. Then, we can determine the throughput of the IUs, that of the D2D pairs, and the harvested energy of EUs as follows. 

The throughput of IU $d_i$ (nats/s/Hz) during the data transfer period of $\ti$ is given by
\begin{align} \label{ati1}
\ti R_{\ti, d_i} (\bw, \bp, \brcv),
\end{align}
where $R_{\ti, d_i} (\bw, \bp, \brcv) = \ds\ln \big{(}1+\frac{|\bh_{(B,d_i)}(\brcv) \bw_{d_i}|^2}{\psi_{\ti, d_i} (\bw, \bp, \brcv)} \big{)}$, and  $\psi_{\ti, d_i} (\bw, \bp, \brcv) = \sum_{l \neq i, d_l \in \clU_I} |\bh_{(B,d_i)}(\brcv) \bw_{d_l}|^2 + \sum_{k=1}^{K} p_k |g_{(k,d_i)}(\brcv)|^2 + \noise_{d_i}$.

Let $\rho\in[0,1]$ denote the energy conversion efficiency of the EUs. The amount of harvested energy of EU $e_j$ during period $\te$ is given by
\begin{align} \label{ae1}
\te \rho E_{\te, e_j} (\bv, \bp, \brcv),
\end{align}
where $E_{\te, e_j} (\bv, \bp, \brcv) = \ds\sum_{l=1}^{U_E} |\bh_{(B,e_j)}(\brcv) \bv_{e_l}|^2 + \ds\sum_{k=1}^{K} p_k |g_{(k,e_j)}(\brcv)|^2$.

Note that in \eqref{ae1}, the energy harvested from the background noise is ignored since the energy is very small compared with that from the beamforming and interference signals.

The throughput achieved by each D2D pair is the total throughput achieved by the D2D pair in both $\ti$ and $\te$. Thus, the throughput of D2D pair $k$ (nats/s/Hz) is determined by
\begin{align} \label{atd1}
\ti R_{\ti,k}(\bw, \bp, \brcv) + \te R_{\te,k}(\bv, \bp, \brcv), 
\end{align}
where $R_{\ti,k}(\bw, \bp, \brcv)= \ds\ln(1+ \frac{ p_k |h_{k}(\brcv)|^2}{\psi_{\ti, k} (\bw, \bp, \brcv)})$ and $R_{\te,k}(\bv, \bp, \brcv)= \ds\ln(1+ \frac{ p_k |h_{k}(\brcv)|^2}{\psi_{\te, k} (\bv, \bp, \brcv)})$. Here, $\psi_{\ti, k} (\bw, \bp, \brcv) $ and $\psi_{\te, k} (\bv, \bp, \brcv) $ are defined as follows:
\[
\begin{array}{c}
\psi_{\ti, k} (\bw, \bp, \brcv) = \sum_{l \in \clK \setminus \{k\}} p_l |g_{l,k}(\brcv)|^2 + \sum_{d_i \in \clU_I} |\bg_{(B,k)}(\brcv) \bw_{d_i}|^2 + \noise_{k},  \\
\psi_{\te, k} (\bv, \bp, \brcv) = \sum_{l \in \clK \setminus \{k\}} p_l |g_{l,k}(\brcv)|^2 + \sum_{e_j \in \clU_E} |\bg_{(B,k)}(\brcv) \bv_{e_j}|^2 + \noise_{k}.
\end{array}
\]

\subsection{Problem Formulation}
In this work, we aim to maximize the worst rate of the IUs, subject to the harvested energy requirements of the EUs and D2D rate thresholds. To achieve this goal, we optimize the i) information beamformer $\bw$, ii) energy beamformer $\bv$, iii) transmit power of D2D transmitters $\bp$, iv) factions of time used for the data transmission and energy transfer of the BS $\bt$, and v) reflection coefficient $\brcv$. In particular, we investigate the max-min throughput optimization problem as follows:
\begin{subequations} \label{op1}
	\begin{align}
	&\ds\max_{\bw, \bv, \bp, \bt=(\ti,\te)\in \mathbb{R}_{+}^2, \brcv} f(\bw, \bv, \bp, \bt, \brcv) \triangleq  \min_{d_i \in \clU_I} \ti R_{\ti, d_i} (\bw, \bp, \brcv) \label{op1a} \\
	\text{s.t. } \; &\eqref{umc}, \nonumber \\
	& \te \rho E_{\te, e_j} (\bv, \bp, \brcv) \geq e_{min}, \forall e_j \in \clU_E , \label{op1b} \\
	&\ti R_{\ti,k}(\bw, \bp, \brcv) + \te R_{\te,k}(\bv, \bp, \brcv) \geq R_{k,min}, \forall k \in \clK, \label{op1c} \\
	& \ti + \te \leq 1, \label{op1d} \\
	&\ti \sum_{d_i \in \clU_I} \|\bw_{d_i}\|^2 + \te \sum_{e_j \in \clU_E} \|\bv_{e_j}\|^2 \leq P_{B, max}, \label{op1e} \\
	&\|\bw_{d_i}\|^2  \leq P_{B, max}; \|\bv_{e_j}\|^2 \leq P_{B, max}, \label{op1f} \\
	&p_k \leq P_{k,max}, \forall k \in \clK, \label{op1g} 
	\end{align}
\end{subequations}
where $e_{min}$ is the energy threshold of the EUs, $R_{k,min}$ is the D2D rate threshold, $P_{k,max}$ and $P_{B, max}$ are the power budgets of the D2D transmitter of pair $k$ and BS, respectively. The constraint in \eqref{umc} is the uni-modulus condition of the IRS elements. The constraint in \eqref{op1b} is the minimum harvesting energy requirement of each EU, and the constraint in \eqref{op1c} is the minimum data rates of the D2D pairs. The constraints in \eqref{op1e}, \eqref{op1f}, and  \eqref{op1g} represents the transmit power constraints of the BS and the D2D pairs. It can be observed from \eqref{op1} that the objective function given in \eqref{op1a} and the constraints in \eqref{umc}, \eqref{op1b}, \eqref{op1c} and \eqref{op1e} are nonconvex. Thus, the optimization problem in \eqref{op1} is nonconvex and difficult to handle due to the time splitting variables and phase shift optimization. To develop the path-following algorithm that improves its feasible value in each iteration, we introduce an inner convex approximation of the nonconvex objective and these nonconvex contraints, which is presented in the next section.

\section{Path-following Algorithm}
\label{path-follo-prob1}
We first introduce two new variables $\tni$ and $\tne$ that are defined as $\tni = 1/\ti$ and $\tne=1/\te$. Then, the optimization problem in \eqref{op1} is reformulated by
\begin{subequations} \label{op2}
	\begin{align}
	&\ds\max_{\bw, \bv, \bp, \btau=(\tni,\tne)\in \mathbb{R}_{+}^2, \brcv} f(\bw, \bv, \bp, \btau, \brcv) \triangleq  \min_{d_i \in \clU_I} (1/\tni) R_{\tni, d_i} (\bw, \bp, \brcv) \label{op2a} \\
	\text{s.t. }  \;  &\eqref{umc}, \nonumber \\
	& (1/\tne) \rho E_{\tne, e_j} (\bv, \bp, \brcv) \geq e_{min}, \forall e_j \in \clU_E , \label{op2b} \\
	& (1/\tni)R_{\tni,k}(\bw, \bp, \brcv) + (1/\tne) R_{\tne,k}(\bv, \bp, \brcv) \geq R_{k,min}, \forall k \in \clK, \label{op2c} \\
	& 1/\tni + 1/\tne \leq 1, \label{op2d} \\
	& (1/\tni) \sum_{d_i \in \clU_I} \|\bw_{d_i}\|^2 + (1/\tne) \sum_{e_j \in \clU_E} \|\bv_{e_j}\|^2 \leq P_{B, max}, \label{op2e} \\
	&\|\bw_{d_i}\|^2  \leq P_{B, max}, \|\bv_{e_j}\|^2 \leq P_{B, max}, \label{op2f} \\
	&p_k \leq P_{k,max}, \forall k \in \clK. \label{op2g} 
	\end{align}
\end{subequations} 

As mentioned earlier, the constraint in \eqref{umc} is nonconvex, and thus before solving the optimization problem, we relax this constraint. In particular, from \eqref{umc}, we have the equivalent condition as $\sum_{n=1}^{N} |\theta_n|^2 = N$. Indeed, for $\brcv = \{(|\theta_n| \in (0,1)), \forall n \in \clN\}$, it follows that $|\theta_n|^2 \leq |\theta_n|, \forall n \in \clN$ and $|\theta_n|^2 = |\theta_n|$ when the constraint in \eqref{umc} is hold. For $|\theta_n|^2 \leq |\theta_n|, \forall n \in \clN$, it is true that $\sum_{n=1}^{N} |\theta_n|^2 \leq \sum_{n=1}^{N} |\theta_n|$. We thus obtain the following inequality 
\begin{align} \label{bie1}
\frac{1}{\sum_{n=1}^{N} |\theta_n|^2} \geq \frac{1}{\sum_{n=1}^{N} |\theta_n|} = \frac{1}{N}.
\end{align} 
The equality condition in \eqref{bie1} holds when the constraint in \eqref{umc} is satisfied. Based on \eqref{bie1}, we define a variable of $\Omega(\brcv)$ as follows: 
\begin{align}
\Omega(\brcv) = \frac{1}{N} - \frac{1}{\sum_{n=1}^{N}  |\theta_n|^2}.
\label{penalty_var}
\end{align}
$\Omega(\brcv)$ measures the degree of satisfaction of the constraint in \eqref{umc1} in the sense that $\Omega(\brcv) \leq 0$ for $|\theta_n| \leq 1$ and $\Omega(\brcv) = 0$ if and only if $\brcv = \{(|\theta_n|=1), \forall n \in \clN\}$. Based on the above manipulations, we can relax the constraint in \eqref{umc} as follows:
\begin{align} \label{umc1}
|\theta_n|^2 \leq 1, \forall n \in \clN. 
\end{align}
Now, we incorporate $\Omega(\brcv)$ given in \eqref{penalty_var} as a penalty function into the objective function in \eqref{op2a}, and we use the constraint in \eqref{umc1} instead of the constraint in \eqref{umc}. The optimization problem in \eqref{op2} is reformulated as follows:
\begin{subequations} \label{op2pe}
	\begin{align}
	&\ds\max_{\bw, \bv, \bp, \btau=(\tni,\tne)\in \mathbb{R}_{+}^2, \brcv} f(\bw, \bv, \bp, \btau, \brcv)  \triangleq  \min_{d_i \in \clU_I} (1/\tni) R_{\tni, d_i} (\bw, \bp, \brcv) + \eta \Omega(\brcv) \label{op2pea} \\
	\text{s.t. } \;  &\eqref{umc1}, \eqref{op2b}-\eqref{op2g},
	\end{align}
\end{subequations} 
where $\eta$ is the penalty parameter.

The optimization problem given in \eqref{op2pe} is still nonconvex because the objective function in \eqref{op2pea} and the constraints in \eqref{op1b} and \eqref{op1c} are nonconvex. Solving this problem is still challenging, especially dealing with the phase shift optimization. Nevertheless, we observe that by fixing either $(\bw, \bv, \bp, \btau=(\tni,\tne))$ or phase shifts $\brcv$, the resulting problem \eqref{op2pe} can be solved efficiently. Therefore, we propose to use an alternating optimization technique~\cite{csiszar1984information} to solve the problem given in \eqref{op2pe}. In particular, we divide the problem \eqref{op2pe} into two sub-problems that are alternatively optimized at each round, i.e., iteration. The first sub-problem, i.e., sub-problem 1, aims to optimize $(\bw, \bv, \bp, \btau=(\tni,\tne))$, and the second sub-problem, i.e., sub-problem 2, aims to optimize $\brcv$.

Let $(\wk, \vk, \pk, \tauk, \rcvk)$ be a feasible point of \eqref{op2pe} that is found from the $(\kappa-1)$-th iteration. In iteration $\kappa$, we fix $\brcv=\rcvk$ and determine $\wkk, \vkk, \pkk$, and $\taukk$, then we fix $\bw=\wkk, \bv=\vkk, \bp=\pkk$, and $\btau=\taukk$ to determine $\rcvkk$.

\subsection{Sub-problem 1}
\label{sec:sub-prob_1_ota}
In this sub-problem, we fix $\rcvk$ and we have the following optimization problem:
\begin{subequations} \label{op2-s1}
	\begin{align}
	&\ds\max_{\bw, \bv, \bp, \btau=(\tni,\tne)\in \mathbb{R}_{+}^2} f_{sub-1}(\bw, \bv, \bp, \btau) \triangleq  \min_{d_i \in \clU_I} (1/\tni) R_{\tni, d_i} (\bw, \bp, \rcvk) \\
	\text{s.t. } \;  &\eqref{op2d}-\eqref{op2g}, \notag \\
	& (1/\tne) \rho E_{\tne, e_j} (\bv, \bp, \rcvk) \geq e_{min}, \forall e_j \in \clU_E , \label{op2b-s1} \\
	& (1/\tni)R_{\tni,k}(\bw, \bp, \rcvk) + (1/\tne) R_{\tne,k}(\bv, \bp, \rcvk) \geq R_{k,min}, \forall k \in \clK. \label{op2c-s1}
	\end{align}
\end{subequations} 

To solve the sub-problem given in \eqref{op2-s1}, we convert the nonconvex objective function and the nonconvex constraints in (\ref{op2b-s1}) and in (\ref{op2c-s1}) to the convex ones. For the objective function, we apply the inequalities \eqref{ie2} in the Appendix to $(1/\tni) R_{\tni, d_i} (\bw, \bp, \rcvk)$ with $x = |\bh_{(B,d_i)}(\rcvk) \bw_{d_i}|^2$, $y = \psi_{\ti, d_i} (\bw, \bp, \rcvk), t= \tni$, and $\bar{x} = |\bh_{(B,d_i)}(\rcvk) \wk_{d_i}|^2, \bar{y} = \psi_{\ti, d_i} (\wk, \pk, \rcvk)$, $\bar{t}= \tnik$. Then, we have
\begin{align}\label{obj-op2-s1}
(1/\tni) R_{\tni, d_i} (\bw, \bp, \rcvk)&\geq a_{\tni, d_i}^{(\kappa)} + b_{\tni, d_i}^{(\kappa)} \Big{(}2-\frac{|\bh_{(B,d_i)}(\rcvk) \wk_{d_i}|^2}{|\bh_{(B,d_i)}(\rcvk) \bw_{d_i}|^2} - \frac{\psi_{\ti, d_i} (\bw, \bp, \rcvk)}{\psi_{\ti, d_i} (\wk, \pk, \rcvk)}\Big{)} - c_{\tni, d_i}^{(\kappa)} \tni \nonumber \\
&\geq a_{\tni, d_i}^{(\kappa)} + b_{\tni, d_i}^{(\kappa)} \Big{(}2-\frac{|\bh_{(B,d_i)}(\rcvk) \wk_{d_i}|^2}{2 \Re \{(\bh_{(B,d_i)}(\rcvk) \bw_{d_i})(\bh_{(B,d_i)}(\rcvk) \wk_{d_i})^{*}\} - |\bh_{(B,d_i)}(\rcvk) \wk_{d_i}|^2} \nonumber \\
&- \frac{\psi_{\ti, d_i} (\bw, \bp, \rcvk)}{\psi_{\ti, d_i} (\wk, \pk, \rcvk)}\Big{)} - c_{\tni, d_i}^{(\kappa)} \tni \nonumber \\
& \triangleq R_{\tni, d_i}^{(\kappa)} ( \bw, \bp, \rcvk),
\end{align}
over the trust region $2 \Re \{(\bh_{(B,d_i)}(\rcvk) \bw_{d_i})(\bh_{(B,d_i)}(\rcvk) \wk_{d_i})^{*}\} - |\bh_{(B,d_i)}(\rcvk) \wk_{d_i}|^2 \geq 0$ for
\begin{eqnarray}
\ds 0 < &a_{\tni, d_i}^{(\kappa)}=& \frac{2}{\tnik} \ln(1+\frac{|\bh_{(B,d_i)}(\rcvk) \wk_{d_i}|^2}{\psi_{\ti, d_i} (\wk, \pk, \rcvk)}), \nonumber\\
\ds 0 < &b_{\tni, d_i}^{(\kappa)}=&  \frac{|\bh_{(B,d_i)}(\rcvk) \wk_{d_i}|^2 / \psi_{\ti, d_i} (\wk, \pk, \rcvk)} {\tnik(1+|\bh_{(B,d_i)}(\rcvk) \wk_{d_i}|^2 / \psi_{\ti, d_i} (\wk, \pk, \rcvk))}, \nonumber \\
\ds 0 < &c_{\tni, d_i}^{(\kappa)}=& \frac{\ln(1+|\bh_{(B,d_i)}(\rcvk) \wk_{d_i}|^2 / \psi_{\ti, d_i} (\wk, \pk, \rcvk))}{(\tnik)^2}. \nonumber
\end{eqnarray}
Function $R_{\tni, d_i}^{(\kappa)} (\bw, \bp, \rcvk)$ is now concave. Next, we consider the constraint in \eqref{op2b-s1}. We use the popular inequality given by
\begin{align}\label{ieq1}
x^2 \geq 2x\bar{x}-\bar{x}^2, \forall x>0, \bar{x}>0.
\end{align}
Then, the nonconvex constraint in \eqref{op2b-s1} can be replaced by the following convex constraint
\begin{align}
E_{\tne, e_j} (\bv, \bp, \rcvk)
&\geq \ds\sum_{l=1}^{U_E} \left[2 \Re \{(\bh_{(B,e_j)}(\rcvk) \bv_{e_l})(\bh_{(B,e_j)}(\rcvk) \vk_{e_l})^{*}\} - |\bh_{(B,e_j)}(\rcvk) \vk_{e_l}|^2 \right] + \ds\sum_{k=1}^{K} p_k |g_{(k,e_j)}(\rcvk)|^2 \nonumber \\
&\triangleq E_{\tne, e_j}^{(\kappa)} (\bv, \bp, \rcvk)  \geq e_{min} \tne / \rho. \label{op3b-s1}
\end{align}

For the nonconstraint in \eqref{op2c-s1}, we apply the inequalities in \eqref{ie2} in the Appendix to $(1/\tni)  R_{\tni,k}(\bw, \bp, \rcvk)$ with $x = p_k |h_{k}(\rcvk)|^2, y = \psi_{\ti, k} (\bw, \bp, \rcvk), t= \tni$ and $\bar{x} = \pk |h_{k}(\rcvk)|^2, \bar{y} =\psi_{\ti, k} (\wk, \pk, \rcvk), \bar{t}= \tnik$. Then, we have
\begin{align}\label{op2c-1}
(1/\tni) R_{\tni,k}(\bw, \bp, \rcvk) 
&\geq a_{\tni, k}^{(\kappa)} + b_{\tni, k}^{(\kappa)} (2-\frac{\pk_k}{p_k} - \frac{\psi_{\ti, k} (\bw, \bp, \rcvk)}{\psi_{\ti, k} (\wk, \pk, \rcvk)}) - c_{\tni, k}^{(\kappa)} \tni \nonumber \\
& \triangleq R_{\tni, k}^{(\kappa)} (\bw, \bp, \rcvk),
\end{align}
where
\begin{eqnarray}
\ds 0 < &a_{\tni,k}^{(\kappa)}=& \frac{2}{\tnik} \ln(1+\frac{\pk_k |h_{k}(\rcvk)|^2}{\psi_{\ti, k} (\wk, \pk, \rcvk)}), \nonumber\\
\ds 0 < &b_{\tni,k}^{(\kappa)}=&  \frac{\pk_k |h_{k}(\rcvk)|^2 / \psi_{\ti, k} (\wk, \pk, \rcvk)} {\tnik(1+\pk |h_{k}(\rcvk)|^2 / \psi_{\ti, k} (\wk, \pk, \rcvk))}, \nonumber \\
\ds 0 < &c_{\tni,k}^{(\kappa)}=& \frac{\ln(1+\pk_k |h_{k}(\rcvk)|^2 / \psi_{\ti, k} (\wk, \pk, \rcvk))}{(\tnik)^2}. \nonumber
\end{eqnarray}
Similarly, we apply the inequalities in \eqref{ie2} in the Appendix to $(1/\tne) R_{\tne,k}(\bv, \bp, \rcvk)$ with $x = p_k |h_{k}(\rcvk)|^2, y = \psi_{\te, k} (\bv, \bp, \rcvk), t= \tni$, and $\bar{x} = \pk_k |h_{k}(\rcvk)|^2, \bar{y} =\psi_{\te, k} (\vk, \pk, \rcvk), \bar{t}= \tnek$. Then, we have
\begin{align}\label{op2c-2}
(1/\tne) R_{\tne,k}(\bv, \bp, \rcvk) 
&\geq a_{\tne, k}^{(\kappa)} + b_{\tne, k}^{(\kappa)} \Big{(}2-\frac{\pk_k}{p_k} - \frac{\psi_{\te, k} (\bv, \bp, \rcvk)}{\psi_{\te, k} (\vk, \pk, \rcvk)}\Big{)} - c_{\tne, k}^{(\kappa)} \tne \nonumber \\
& \triangleq R_{\tne, k}^{(\kappa)} (\bv, \bp, \rcvk),
\end{align}
where
\begin{eqnarray}
\ds 0 < &a_{\tne,k}^{(\kappa)}=& \frac{2}{\tnek} \ln(1+\frac{\pk_k |h_{k}(\rcvk)|^2}{\psi_{\te, k} (\vk, \pk, \rcvk)}), \nonumber\\
\ds 0 < &b_{\tne,k}^{(\kappa)}=&  \frac{\pk_k |h_{k}(\rcvk)|^2 / \psi_{\te, k} (\vk, \pk, \rcvk)} {\tnek(1+\pk_k |h_{k}(\rcvk)|^2 / \psi_{\te, k} (\vk, \pk, \rcvk))}, \nonumber \\
\ds 0 < &c_{\tne,k}^{(\kappa)}=& \frac{\ln(1+\pk_k |h_{k}(\rcvk)|^2 / \psi_{\te, k} (\vk, \pk, \rcvk))}{(\tnek)^2}.\nonumber
\end{eqnarray}
Based on \eqref{op2c-1} and \eqref{op2c-2}, the nonconvex constraint in \eqref{op2c-s1} is innerly approximated by the following convex constraint
\begin{align}\label{op3c-s1} 
R_{\tni, k}^{(\kappa)} (\bw, \bp, \rcvk) + R_{\tne, k}^{(\kappa)} (\bv, \bp, \rcvk) \geq R_{k,min}, \forall k \in \clK.
\end{align}
Given \eqref{obj-op2-s1}, \eqref{op3b-s1} and \eqref{op3c-s1}, the sub-problem 1 can be expressed as
\begin{subequations} \label{op3-s1}
	\begin{align}
	&\ds\max_{\bw, \bv, \bp, \btau=(\tni,\tne)\in \mathbb{R}_{+}^2} f_{sub-1}^{(\kappa)}(\bw, \bv, \bp, \btau, \rcvk)  \nonumber \\
	&\triangleq  \min_{d_i \in \clU_I} R_{\tni, d_i}^{(\kappa)} (\bw, \bp, \rcvk) + \eta \Omega(\rcvk) \label{op3a-s1} \\
	\text{s.t. } \;  &\eqref{op2d}-\eqref{op2g}, \eqref{op3b-s1} \text{ and } \eqref{op3c-s1}. \nonumber
	\end{align}
\end{subequations} 
Function $f_{sub-1}^{(\kappa)}(\bw, \bv, \bp, \btau, \rcvk)$ is concave since the first term in \eqref{op3a-s1} is concave (i.e., the minimum of concave function~\cite{Tuybook}) and the second term is concave. Thus, instead of \eqref{op2-s1}, we solve the convex optimization problem \eqref{op3-s1}, which also generates $(\wkk, \vkk, \pkk, \taukk)$ in the next iteration. The computational complexity of the algorithm to solve the convex problem in \eqref{op3-s1} is \cite{peaucelle2002user}
\begin{align}\label{com_complexity}
\clO(\alpha^2\beta^{2.5}+\beta^{3.5}),
\end{align}
where $\alpha=2(M+1)+K$ that is the number of the decision variables and $\beta=U_I+2(U_E+K+1)$ that is the number of the constraints. Note that $(\wk, \vk, \pk, \tauk)$ is a feasible point to \eqref{op3-s1}, meaning that $f_{sub-1}^{(\kappa)}(\wk, \vk, \pk, \tauk, \rcvk)=f_{sub-1}(\wk, \vk, \pk, \tauk)$. Meanwhile, $(\wkk, \vkk, \pkk, \taukk)$ is the optimal solution of \eqref{op3-s1}, and thus we have
\begin{align}
f_{sub-1}^{(\kappa)}(\wkk, \vkk, \pkk, \taukk, \rcvk)>f_{sub-1}^{(\kappa)}(\wk, \vk, \pk, \tauk, \rcvk),
\end{align}
for $(\wkk, \vkk, \pkk, \taukk) \neq (\wk, \vk, \pk, \tauk)$. Consider again \eqref{obj-op2-s1}, we have
\begin{align}
f(\wk, \vk, \pk, \tauk, \rcvk)&=f_{sub-1}^{(\kappa)}(\wk, \vk, \pk, \tauk, \rcvk) \\
&< f_{sub-1}^{(\kappa)}(\wkk, \vkk, \pkk, \taukk, \rcvk) \\
&\leq f(\wkk, \vkk, \pkk, \taukk, \rcvk).
\end{align}
This shows that the optimal solution $(\wkk, \vkk, \pkk, \taukk)$ of \eqref{op3-s1} satisfies the convergence condition: $f(\wkk, \vkk, \pkk, \taukk, \rcvk) > f(\wk, \vk, \pk, \tauk, \rcvk).$

\subsection{Sub-problem 2}
We fix variables $\wk, \vk, \pk, \tauk$ and optimize $\brcv$. Thus, we have the following problem:
\begin{subequations} \label{op2-s2}
	\begin{align}
	&\ds\max_{\brcv} f_{sub-2}(\wk, \vk, \pk, \tauk, \brcv)  \triangleq  \min_{d_i \in \clU_I} (1/\tnik) R_{\tni, d_i} (\wk, \pk, \brcv) + \eta \Omega(\brcv) \label{op2a-s2} \\
	\text{s.t. } \;  &\eqref{umc1}, \nonumber \\
	& (1/\tnek) \rho E_{\tne, e_j} (\vk, \pk, \brcv) \geq e_{min}, \forall e_j \in \clU_E , \label{op2b-s2} \\
	& (1/\tnik)R_{\tni,k}(\wk, \pk, \brcv) + (1/\tnek) R_{\tne,k}(\vk, \pk, \brcv) \geq R_{k,min}, \forall k \in \clK. \label{op2c-s2} 
	\end{align}
\end{subequations} 

Similar to Section~\ref{sec:sub-prob_1_ota}, to solve the problem in \eqref{op2-s2}, we convert the nonconvex objective function in \eqref{op2a-s2}, and the nonconvex constraints in \eqref{op2b-s2} and \eqref{op2c-s2} to the convex ones. For this, we first rewrite $\bh_{(B,d_i)}(\brcv)$, $\bh_{(B,e_j)}(\brcv)$, $g_{(k,d_i)}(\brcv)$, $g_{(k,e_j)}(\brcv)$, $h_{k}(\brcv)$, $g_{l,k}(\brcv)$ and $\bg_{(B,k)}(\brcv)$ as $\bh_{(B,d_i)}(\brcv) = \bh_{(B,d_i)}^H + \brcv \text{diag}(\bh_{(r,d_i)}^H) \bG$, $\bh_{(B,e_j)}(\brcv) = \bh_{(B,e_j)}^H + \brcv \text{diag}(\bh_{(r,e_j)}^H) \bG$, $g_{(k,d_i)}(\brcv) = g_{(k,d_i)} + \brcv \text{diag}(\bh_{(r,d_i)}^H) \bh_{k,r}$, $g_{(k,e_j)}(\brcv) = g_{(k,e_j)} + \brcv \text{diag}(\bh_{(r,e_j)}^H) \bh_{k,r}$, $h_{k}(\brcv) = h_{k} + \brcv \text{diag}(\bh_{(r,k)}^H) \bh_{k,r}$, $g_{l,k}(\brcv) = g_{l,k} + \brcv \text{diag}(\bh_{(r,k)}^H) \bh_{l,r}$ and $\bg_{(B,k)}(\brcv) = \bg_{(B,k)}^H + \brcv \text{diag}(\bh_{(r,k)}^H) \bG$. Then, we convert the noncovex objective function and the nonconvex constraints in~\eqref{op2-s2} as follows.

We apply inequalities in \eqref{ie1} in the Appendix to the term $R_{\tni, d_i} (\wk, \pk, \brcv)$ in \eqref{op2a-s2} with $x = |\bh_{(B,d_i)}(\brcv) \wk_{d_i}|^2, y = \psi_{\ti, d_i} (\wk, \pk, \brcv)$, $\bar{x} = |\bh_{(B,d_i)}(\rcvk) \wk_{d_i}|^2$, and $\bar{y} = \psi_{\ti, d_i} (\wk, \pk, \rcvk)$. Then, we have
\begin{align}\label{op3-ob}
&R_{\tni, d_i} (\wk, \pk, \brcv) \nonumber \\
&\geq a_{\tni, d_i}^{(\kappa)} + b_{\tni, d_i}^{(\kappa)}\Big{(}2-\frac{|\bh_{(B,d_i)}(\rcvk) \wk_{d_i}|^2}{|\bh_{(B,d_i)}(\brcv) \wk_{d_i}|^2} - \frac{\psi_{\ti, d_i} (\wk, \pk, \brcv)} {\psi_{\ti, d_i} (\wk, \pk, \rcvk)}\Big{)} \nonumber \\
&\geq a_{\tni, d_i}^{(\kappa)} + b_{\tni, d_i}^{(\kappa)} \Big{(}2-\frac{|\bh_{(B,d_i)}(\rcvk) \wk_{d_i}|^2}{2 \Re \{(\bh_{(B,d_i)}(\brcv) \wk_{d_i})(\bh_{(B,d_i)}(\rcvk) \wk_{d_i})^{*}\} - |\bh_{(B,d_i)}(\rcvk) \wk_{d_i}|^2} - \frac{\psi_{\ti, d_i} (\wk, \pk, \brcv)}{\psi_{\ti, d_i} (\wk, \pk, \rcvk)}\Big{)} \nonumber \\
& \triangleq R_{\tni, d_i}^{(\kappa)} (\wk, \pk, \brcv) 
\end{align}
over the trust region $2 \Re \{(\bh_{(B,d_i)}(\brcv) \wk_{d_i})(\bh_{(B,d_i)}(\rcvk) \wk_{d_i})^{*}\} - |\bh_{(B,d_i)}(\rcvk) \wk_{d_i}|^2 \geq 0,$
where
\begin{eqnarray}
\ds 0 < &a_{\tni, d_i}^{(\kappa)}=& \ln(1+\frac{|\bh_{(B,d_i)}(\rcvk) \wk_{d_i}|^2}{\psi_{\ti, d_i} (\wk, \pk, \rcvk)}), \nonumber\\
\ds 0 < &b_{\tni, d_i}^{(\kappa)}=&  \frac{|\bh_{(B,d_i)}(\rcvk) \wk_{d_i}|^2 / \psi_{\ti, d_i} (\wk, \pk, \rcvk)} {1+|\bh_{(B,d_i)}(\rcvk) \wk_{d_i}|^2 / \psi_{\ti, d_i} (\wk, \pk, \rcvk)}. \nonumber
\end{eqnarray}
Moreover, using the inequality given in \eqref{ieq1}, we have
\begin{align}\label{op3-ob1}
\Omega(\brcv) &\geq  \frac{1}{N} - \frac{1}{\sum_{n=1}^{N}  (2 \Re\{(\rcvk_n)^{*} \theta_n\}  - |\rcvk_n|^2)} \triangleq \Omega^{(\kappa)}(\brcv) 
\end{align}
over the trust region $\sum_{n=1}^{N}  (2 \Re\{(\rcvk_n)^{*} \theta_n\}  - |\rcvk_n|^2) \geq 0$. From \eqref{op3-ob} and \eqref{op3-ob1}, we have
\begin{align}\label{op3-obj}
f_{sub-2}(\wk, \vk, \pk, \tauk, \brcv) &\geq f_{sub-2}^{(\kappa)}(\wk, \vk, \pk, \tauk, \brcv)  \nonumber \\
	&\triangleq  \min_{d_i \in \clU_I} (1/\tnik) R_{\tni, d_i}^{(\kappa)} (\wk, \pk, \brcv)  + \eta \Omega^{(\kappa)}(\brcv)
\end{align}
where $f_{sub-2}^{(\kappa)}(\wk, \vk, \pk, \tauk, \brcv)$ is now concave since the first term in \eqref{op3-obj} is concave, i.e., a type of minimum of concave function \cite{Tuybook}, and the second term is already concave. 

For the constraint in \eqref{op2b-s2}, by using the inequality \eqref{op2b-s2}, the nonconvex constraint can be replaced by the following convex constraint:
\begin{align}\label{op3b-s2}
E_{\tne, e_j} (\vk, \pk, \brcv)
&\geq \ds\sum_{l=1}^{U_E} \left[2 \Re \{(\bh_{(B,e_j)}(\brcv) \vk_{e_l})(\bh_{(B,e_j)}(\rcvk) \vk_{e_l})^{*}\} - |\bh_{(B,e_j)}(\rcvk) \vk_{e_l}|^2 \right] +  \nonumber \\
&+ \ds\sum_{k=1}^{K} \pk \left[2 \Re \{g_{(k,e_j)}(\brcv)g_{(k,e_j)}(\rcvk)^{*}\} - |g_{(k,e_j)}(\rcvk)|^2 \right] \nonumber \\ %
&\triangleq E_{\tne, e_j}^{(\kappa)} (\vk, \pk, \brcv)  \geq e_{min} \tnek / \rho.
\end{align}

To convert the nonconvex constraint in \eqref{op2c-s2} to the convex constraint, we apply the inequalities in \eqref{ie1} in the Appendix to the term of $R_{\tni,k}(\wk, \pk, \brcv)$ in \eqref{op2c-s2} with $x = \pk |h_{k}(\brcv)|^2, y = \psi_{\ti, k} (\wk, \pk, \brcv)$, and $\bar{x} = \pk |h_{k}(\rcvk)|^2, \bar{y} = \psi_{\ti, k} (\wk, \pk, \rcvk)$. Then, we have
\begin{align}\label{op3-Rti}
R_{\tni,k}(\wk, \pk, \brcv) 
&\geq a_{\tni, k}^{(\kappa)} + b_{\tni, k}^{(\kappa)} \Big{(}2-\frac{|h_{k}(\rcvk)|^2}{|h_{k}(\brcv)|^2}-\frac{\psi_{\ti, k} (\wk, \pk, \brcv)}{\psi_{\ti, k} (\wk, \pk, \rcvk)}\Big{)} \nonumber \\
&\geq a_{\tni, k}^{(\kappa)} + b_{\tni, k}^{(\kappa)} \Big{(}2-\frac{|h_{k}(\rcvk)|^2}{2 \Re \{h_{k}(\brcv)h_{k}(\rcvk)^{*}\} - |h_{k}(\rcvk)|^2}-\frac{\psi_{\ti, k} (\wk, \pk, \brcv)}{\psi_{\ti, k} (\wk, \pk, \rcvk)}\Big{)} \nonumber \\
& \triangleq R_{\tni, k}^{(\kappa)} (\wk, \pk, \brcv) 
\end{align}
over the trust region $2 \Re \{h_{k}(\brcv)h_{k}(\rcvk)^{*}\} - |h_{k}(\rcvk)|^2 \geq 0$, where 
\begin{eqnarray}
\ds 0 < &a_{\tni,k}^{(\kappa)}=& \ln(1+\frac{\pk |h_{k}(\rcvk)|^2}{\psi_{\ti, k} (\wk, \pk, \rcvk)}), \nonumber\\
\ds 0 < &b_{\tni,k}^{(\kappa)}=&  \frac{\pk |h_{k}(\rcvk)|^2 / \psi_{\ti, k} (\wk, \pk, \rcvk)}{1+\pk |h_{k}(\rcvk)|^2 / \psi_{\ti, k} (\wk, \pk, \rcvk)}. \nonumber
\end{eqnarray}
We do the same way with the term of $R_{\tne,k}(\vk, \pk, \brcv)$ in \eqref{op2c-s2} with $x = \pk |h_{k}(\brcv)|^2, y = \psi_{\te, k} (\vk, \pk, \brcv)$, and $\bar{x} = \pk |h_{k}(\rcvk)|^2, \bar{y} = \psi_{\te, k} (\vk, \pk, \rcvk)$. Then, we have
\begin{align}\label{op3-Rte}
R_{\tne,k}(\vk, \pk, \brcv) 
&\geq a_{\tne, k}^{(\kappa)} + b_{\tne, k}^{(\kappa)} \Big{(}2-\frac{|h_{k}(\rcvk)|^2}{|h_{k}(\brcv)|^2}-\frac{\psi_{\te, k} (\vk, \pk, \brcv)}{\psi_{\te, k} (\vk, \pk, \rcvk)}\Big{)} \nonumber \\
&\geq a_{\tne, k}^{(\kappa)} + b_{\tne, k}^{(\kappa)} \Big{(}2-\frac{|h_{k}(\rcvk)|^2}{2 \Re \{h_{k}(\brcv)h_{k}(\rcvk)^{*}\} - |h_{k}(\rcvk)|^2}-\frac{\psi_{\te, k} (\vk, \pk, \brcv)}{\psi_{\te, k} (\vk, \pk, \rcvk)}\Big{)} \nonumber \\
& \triangleq R_{\tne, k}^{(\kappa)} (\vk, \pk, \brcv),
\end{align}
where
\begin{eqnarray}
\ds 0 < &a_{\tne,k}^{(\kappa)}=& \ln(1+\frac{\pk |h_{k}(\rcvk)|^2}{\psi_{\te, k} (\vk, \pk, \rcvk)}), \nonumber\\
\ds 0 < &b_{\tne,k}^{(\kappa)}=&  \frac{\pk |h_{k}(\rcvk)|^2 / \psi_{\te, k} (\vk, \pk, \rcvk)}{1+\pk |h_{k}(\rcvk)|^2 / \psi_{\te, k} (\vk, \pk, \rcvk)}. \nonumber 
\end{eqnarray}
Based on \eqref{op3-Rti} and \eqref{op3-Rte}, the nonconvex constraint in \eqref{op2c-s2} is innerly approximated by the following convex constraint
\begin{align} \label{op3c-s2} 
(1/\tnik)R_{\tni,k}^{(\kappa)}(\wk, \pk, \brcv) + (1/\tnek) R_{\tne,k}^{(\kappa)}(\vk, \pk, \brcv) \geq R_{k,min}, \forall k \in \clK.
\end{align}

Based on \eqref{op3-obj}, \eqref{op3b-s2}, and \eqref{op3c-s2}, instead of \eqref{op2-s2}, we solve the following convex optimization problem
	\begin{align}
	\label{op3-s2}
	&\ds\max_{\brcv} f_{sub-2}^{(\kappa)}(\wk, \vk, \pk, \tauk, \brcv)  \\
	\text{s.t. } \;  &\eqref{umc1}, \eqref{op3b-s2}, \text{ and } \eqref{op3c-s2}. \notag
	 \end{align}
The solution of \eqref{op3-s2} is also used as $\rcvkk$ in the next iteration. The computational complexity of the algorithm to solve the convex problem in $\eqref{op3-s2}$ is $\clO(\alpha^2\beta^{2.5}+\beta^{3.5})$, where $\alpha=N$ and $\beta=N+U_E+K$. The overall algorithm for the max-min throughput optimization problem in \eqref{op2} is shown in Algorithm~\ref{Alg_1}. This algorithm is namely N-OTA algorithm, meaning that there is no orthogonal time allocation between the D2D communications and the integrated data and energy network. To enhance the computational efficiency of the algorithm, it is important to generate a feasible point. Moreover, we need to determine the penalty parameter $\eta$. 

\subsection{Generation of a Feasible Point and Selection of $\eta$}
\label{sec:feasible_gen}
We fix $\tau^{(0)}=(\tni^{(0)}, \tne^{(0)})$, $\theta^{(0)}$ that satisfies the convex constraints in \eqref{umc} and \eqref{op2d}, and we randomly generate $(w^{(0)},v^{(0)},p^{(0)})$ feasible for \eqref{op2d}-\eqref{op2g}. Then, we solve the following problem
\begin{subequations} \label{init1}
	\begin{align}
	&\ds\max_{\bw, \bv, \bp} \mu  \label{init1a} \\
	\text{s.t. } \;  &\eqref{op2f}-\eqref{op2g}, \nonumber \\
	& (1/\tni^{(0)}) R_{\tni, d_i}^{(\kappa)} (\bw, \bp, \rcvk) \geq \mu, \forall d_i \in \clU_I, \label{init1b} \\
	& E_{\tne, e_j}^{(\kappa)} (\bv, \bp, \rcvk)  \geq e_{min} \tne^{(0)} \mu / \rho, \forall e_j \in \clU_E, \label{init1c} \\
	& R_{\tni, k}^{(\kappa)} (\bw, \bp, \rcvk) + R_{\tne, k}^{(\kappa)} (\bv, \bp, \rcvk) \geq R_{k,min} \mu, \forall k \in \clK, \label{init1d} \\
	& (1/\tni^{(0)}) \sum_{d_i \in \clU_I} \|\bw_{d_i}\|^2 + (1/\tne^{(0)}) \sum_{e_j \in \clU_E} \|\bv_{e_j}\|^2 \leq P_{B, max}, \label{init1e}
	\end{align}
\end{subequations} 
until $\mu \geq 1$. Solving \eqref{init1} generates a feasible set of $(\wk, \vk, \pk, \tau^{(0)}= (\tni^{(0)},\tne^{(0)}), \theta^{(0)})$ that is considered to be a feasible point for \eqref{op2}. Then, we can select a value of $\eta$ as
\begin{align}\label{eta_1}
\eta = - \left[ \min_{d_i \in \clU_I} (1/\tni^{(0)}) R_{\tni, d_i} (\wk, \pk, \theta^{(0)})\right] / \Omega(\theta^{(0)})
\end{align}
so as to ensure the same magnitude between objective function and penalty function~\cite{shi2017global}.


\begin{algorithm}
\footnotesize
         \caption{\footnotesize N-OTA algorithm for \eqref{op2}}\label{Alg_1}
        \begin{algorithmic}[1]
          \State Initialize: Generate any values of $\tau^{(0)}=(\tni^{(0)},\tne^{(0)})$ and $\theta^{(0)}$ satisfying the convex constraints in \eqref{umc} and \eqref{op2d}, then solve \eqref{init1} to receive a feasible set of $(\wk, \vk, \pk, \tau^{(0)}= (\tni^{(0)},\tne^{(0)}), \theta^{(0)})$ for \eqref{op2}. Set $\kappa = 0$;
	\State Compute $\eta$ according to \eqref{eta_1};
           \Repeat 
		\State Solve the convex problem in \eqref{op3-s1} for $\brcv = \theta^{(\kappa)}$ to generate $(\wkk, \vkk, \pkk, \taukk)$;
		\State Solve the convex problem in \eqref{op3-s2} for $(\bw, \bv, \bp, \btau)=(\wkk, \vkk, \pkk, \taukk)$ to generate $\theta^{(\kappa+1)}$;
		\State $\kappa \gets \kappa+1$;
	\Until Convergence.
	\State Output $(\wk, \vk, \pk, \tauk=(\tni^{(\kappa)}, \tne^{(\kappa)}), \rcvk)$ 
       \end{algorithmic}
    \end{algorithm}
\section{Orthogonal Time Allocation Scenario}
\label{sec:ota}
In Section~\ref{path-follo-prob1}, we consider the IRS-aided DED2D system in which the D2D transmitters perform their communications during the data transfer and energy transfer time. The N-OTA scenario can help to improve the spectrum efficiency, but raises the interference management. In this section, we consider again the IRS-aided DED2D system in which there is an orthogonal time allocation (OTA) among the data transfer phase, energy transfer phase, and D2D communication. In particular, the BS allocates the time fractions to the data transfer, energy transfer, and D2D communication. As such, the N-OTA can be considered to be the OTA in which time fraction allocated to the D2D communication is zero. In the OTA scenario, there are no interference caused by D2D communications to the integrated data and energy network, but the time fractions for the data and energy transfers reduce. Thus, it is interesting to discuss the efficiency of the N-OTA and OTA scenarios, which will be presented in the simulation results.

We first introduce a new variable $\td$ to indicate the time faction allocated to the D2D communications. Then, the throughput at IUs, the harvested energy at EUs, and the throughput at the D2D pairs are determined as follows. The throughput at IU $d_i$ (nats/s/Hz) during time fraction $\ti$ is $\ti \widetilde{R}_{\ti, d_i} (\bw, \brcv)$, where $R_{\ti, d_i} (\bw, \brcv) = \ds\ln(1+\frac{|\bh_{(B,d_i)}(\brcv) \bw_{d_i}|^2}{\widetilde{\psi}_{\ti, d_i} (\bw, \brcv)})$ with $\widetilde{\psi}_{\ti, d_i} (\bw, \brcv) = \sum_{l \neq i, d_l \in \clU_I} |\bh_{(B,d_i)}(\brcv) \bw_{d_l}|^2 + \noise_{d_i}$. The harvested energy during time fraction $\te$ of EU $e_j$ is $\te \rho \widetilde{E}_{\te, e_j} (\bv, \brcv)$, where $\widetilde{E}_{\te, e_j} (\bv, \brcv) = \ds\sum_{l=1}^{U_E} |\bh_{(B,e_j)}(\brcv) \bv_{e_l}|^2$. The throughput at D2D pair $k$ (nats/s/Hz) is $\td \widetilde{R}_{\td,k}(\bp, \brcv)$, where $\widetilde{R}_{\td,k}(\bp, \brcv)= \ds\ln(1+ \frac{ p_k |h_{k}(\brcv)|^2}{\widetilde{\psi}_{\td, k} (\bw, \bp, \brcv)})$ with $\widetilde{\psi}_{\td, k} (\bp, \brcv) = \sum_{l \in \clK \setminus \{k\}} p_l |g_{l,k}(\brcv)|^2 + \noise_{k}$.

We now formulate the optimization problem for the OTA scenario. The problem is similar to \eqref{op1} in which the variable $\td$ that indicates the time fraction allocated to the D2D communications is included in $\bt$, i.e., $\bt=(\ti,\te,\td)$. Mathematically, the optimization problem is formulated as follows:
\begin{subequations} \label{opo1}
	\begin{align}
	&\ds\max_{\bw, \bv, \bp, \bt=(\ti,\te,\td)\in \mathbb{R}_{+}^3, \brcv} f(\bw, \bv, \bp, \bt, \brcv)  \triangleq  \min_{d_i \in \clU_I} \ti \widetilde{R}_{\ti, d_i} (\bw, \brcv) \label{opo1a} \\
	\text{s.t. } \; &\eqref{umc}, \nonumber \\
	& \te \rho \widetilde{E}_{\te, e_j} (\bv, \brcv) \geq e_{min}, \forall e_j \in \clU_E , \label{opo1b} \\
	&\td \widetilde{R}_{\td,k}(\bp, \brcv) \geq R_{k,min}, \forall k \in \clK, \label{opo1c} \\
	& \ti + \te + \td \leq 1, \label{opo1d} \\
	&\ti \sum_{d_i \in \clU_I} \|\bw_{d_i}\|^2 + \te \sum_{e_j \in \clU_E} \|\bv_{e_j}\|^2 \leq P_{B, max}  (1-\td), \label{opo1e} \\
	&\|\bw_{d_i}\|^2  \leq P_{B, max}, \|\bv_{e_j}\|^2 \leq P_{B, max}, \label{opo1f} \\
	&\td p_k \leq P_{k,max}, \forall k \in \clK. \label{opo1g} 
	\end{align}
\end{subequations}

To solve the non-convex problem given in \eqref{opo1}, we introduce new variables $\tni = 1/\ti$, $\tne=1/\te$, and $\tnd=1/\td$. Then, the problem in \eqref{opo1} is equivalently expressed as
\begin{subequations} \label{opo2}
	\begin{align}
	&\ds\max_{\bw, \bv, \bp, \btau=(\tni,\tne, \tnd)\in \mathbb{R}_{+}^3, \brcv} f(\bw, \bv, \bp, \btau, \brcv) \triangleq  \min_{d_i \in \clU_I} (1/\tni) \widetilde{R}_{\tni, d_i} (\bw, \brcv) \label{opo2a} \\
	\text{s.t. } \;  &\eqref{umc}, \nonumber \\
	& (1/\tne) \rho \widetilde{E}_{\tne, e_j} (\bv, \brcv) \geq e_{min}, \forall e_j \in \clU_E , \label{opo2b} \\
	& (1/\tnd)\widetilde{R}_{\tnd,k}(\bp, \brcv) \geq R_{k,min}, \forall k \in \clK, \label{opo2c} \\
	& 1/\tni + 1/\tne + 1/\tnd \leq 1, \label{opo2d} \\
	& (1/\tni) \sum_{d_i \in \clU_I} \|\bw_{d_i}\|^2 + (1/\tne) \sum_{e_j \in \clU_E} \|\bv_{e_j}\|^2 \leq P_{B, max} (1-1/\tnd), \label{opo2e} \\
	&\|\bw_{d_i}\|^2  \leq P_{B, max}, \|\bv_{e_j}\|^2 \leq P_{B, max}, \label{opo2f} \\
	&p_k \leq P_{k,max} \tnd, \forall k \in \clK. \label{opo2g} 
	\end{align}
\end{subequations} 
We further incorporate the penalty function $\Omega(\brcv)$ into the optimization problem in \eqref{opo2} as
\begin{subequations} \label{opo2pe}
	\begin{align}
	&\ds\max_{\bw, \bv, \bp, \btau=(\tni,\tne,\tnd)\in \mathbb{R}_{+}^3, \brcv} f(\bw, \bv, \bp, \btau, \brcv)  \triangleq  \min_{d_i \in \clU_I} (1/\tni) \widetilde{R}_{\tni, d_i} (\bw, \brcv) + \eta \Omega(\brcv) \label{opo2pea} \\
	\text{s.t. } \;  &\eqref{umc1}, \eqref{opo2b}-\eqref{opo2g}.
	\end{align}
\end{subequations} 
Similar to Section~\ref{path-follo-prob1}, we divide the problem in \eqref{opo2pe} into two sub-problems that are alternatively optimized at each round, i.e., iteration. Sub-problem 1 aims to optimize $\bw, \bv, \bp, \btau=(\tni,\tne,\tnd)$, and sub-problem 2 is to optimize $\brcv$.
\subsection{Sub-problem 1}
\label{sub-problem1}
In this sub-problem, we fix $\rcvk$ and solve the following optimization problem:
\begin{subequations} \label{opo2-s1}
	\begin{align}
	&\ds\max_{\bw, \bv, \bp, \btau=(\tni,\tne,\tnd)\in \mathbb{R}_{+}^3} f_{sub-1}(\bw, \bv, \bp, \btau, \rcvk) \triangleq  \min_{d_i \in \clU_I} (1/\tni) \widetilde{R}_{\tni, d_i} (\bw, \rcvk) + \eta \Omega(\rcvk) \label{opo2a-s1} \\
	\text{s.t. } \;  & \eqref{opo2d}-\eqref{opo2g}, \nonumber \\
	& (1/\tne) \rho \widetilde{E}_{\tne, e_j} (\bv, \rcvk) \geq e_{min}, \forall e_j \in \clU_E, \label{opo2b-s1} \\
	& (1/\tnd)\widetilde{R}_{\tnd,k}(\bp, \rcvk)  \geq R_{k,min}, \forall k \in \clK. \label{opo2c-s1} 
	\end{align}
\end{subequations} 
Suppose that $(\wk, \vk, \pk, \tauk)$ is a feasible point for \eqref{opo2-s1} found from iteration $(\kappa-1)$. To solve the problem in \eqref{opo2-s1}, we convert the objective function and constraints in~\eqref{opo2b-s1} and~\eqref{opo2c-s1} to convex ones as follows. We apply the inequalities in \eqref{ie2} in the Appendix to the term of $(1/\tni) \widetilde{R}_{\tni, d_i} (\bw, \rcvk)$ in the objective function with $x = |\bh_{(B,d_i)}(\rcvk) \bw_{d_i}|^2, y = \widetilde{\psi}_{\ti, d_i} (\bw, \rcvk), t= \tni$ and $\bar{x} = |\bh_{(B,d_i)}(\rcvk) \wk_{d_i}|^2, \bar{y} = \widetilde{\psi}_{\ti, d_i} (\wk, \rcvk), \bar{t}= \tnik$. Then, we have 
\begin{align}
&(1/\tni) \widetilde{R}_{\tni, d_i} (\bw, \rcvk) \nonumber \\
&\geq \widetilde{a}_{\tni, d_i}^{(\kappa)} + \widetilde{b}_{\tni, d_i}^{(\kappa)} \Big{(}2-\frac{|\bh_{(B,d_i)}(\rcvk) \wk_{d_i}|^2}{|\bh_{(B,d_i)}(\rcvk) \bw_{d_i}|^2} - \frac{\widetilde{\psi}_{\ti, d_i} (\bw, \rcvk)}{\widetilde{\psi}_{\ti, d_i} (\wk, \rcvk)}\Big{)} - \widetilde{c}_{\tni, d_i}^{(\kappa)} \tni \nonumber \\
&\geq \widetilde{a}_{\tni, d_i}^{(\kappa)} + \widetilde{b}_{\tni, d_i}^{(\kappa)} \Big{(}2-\frac{|\bh_{(B,d_i)}(\rcvk) \wk_{d_i}|^2}{2 \Re \{(\bh_{(B,d_i)}(\rcvk) \bw_{d_i})(\bh_{(B,d_i)}(\rcvk) \wk_{d_i})^{*}\} - |\bh_{(B,d_i)}(\rcvk) \wk_{d_i}|^2} - \frac{\widetilde{\psi}_{\ti, d_i} (\bw,  \rcvk)} {\widetilde{\psi}_{\ti, d_i} (\wk, \rcvk)}\Big{)} - \nonumber \\
& - \widetilde{c}_{\tni, d_i}^{(\kappa)} \tni \nonumber \\
& \triangleq \widetilde{R}_{\tni, d_i}^{(\kappa)} (\bw, \rcvk) 
\end{align}
over the trust region $2 \Re \{(\bh_{(B,d_i)}(\rcvk) \bw_{d_i})(\bh_{(B,d_i)}(\rcvk) \wk_{d_i})^{*}\} - |\bh_{(B,d_i)}(\rcvk) \wk_{d_i}|^2 \geq 0,$ where
\begin{eqnarray}
\ds 0 < &\widetilde{a}_{\tni, d_i}^{(\kappa)}=& \frac{2}{\tnik} \ln(1+\frac{|\bh_{(B,d_i)}(\rcvk) \wk_{d_i}|^2}{\widetilde{\psi}_{\ti, d_i} (\wk,\rcvk)}), \nonumber\\
\ds 0 < &\widetilde{b}_{\tni, d_i}^{(\kappa)}=&  \frac{|\bh_{(B,d_i)}(\rcvk) \wk_{d_i}|^2 / \widetilde{\psi}_{\ti, d_i} (\wk,\rcvk)} {\tnik(1+|\bh_{(B,d_i)}(\rcvk) \wk_{d_i}|^2 / \widetilde{\psi}_{\ti, d_i} (\wk,\rcvk))}, \nonumber \\
\ds 0 < &\widetilde{c}_{\tni, d_i}^{(\kappa)}=& \frac{\ln(1+|\bh_{(B,d_i)}(\rcvk) \wk_{d_i}|^2 / \widetilde{\psi}_{\ti, d_i} (\wk,\rcvk))}{(\tnik)^2}. \nonumber
\end{eqnarray}
Meanwhile, the constraint in \eqref{opo2b-s1} can be replaced by
\begin{align}
\widetilde{E}_{\tne, e_j} (\bv, \rcvk)
&\geq \ds\sum_{l=1}^{U_E} \left[2 \Re \{(\bh_{(B,e_j)}(\rcvk) \bv_{e_l})(\bh_{(B,e_j)}(\rcvk) \vk_{e_l})^{*}\} - |\bh_{(B,e_j)}(\rcvk) \vk_{e_l}|^2 \right] \nonumber \\
&\triangleq \widetilde{E}_{\tne, e_j}^{(\kappa)} (\bv, \rcvk)  \geq e_{min} \tne / \rho.
\end{align}
Similarly, we apply the inequalities in \eqref{ie2} in the Appendix to the term of $(1/\tnd)\widetilde{R}_{\tnd,k}(\bp, \rcvk)$ in \eqref{opo2c-s1} with $x = p_k |h_{k}(\rcvk)|^2, y = \widetilde{\psi}_{\td, k} (\bp, \rcvk), t= \tnd, \bar{x} = \pk |h_{k}(\rcvk)|^2, \bar{y} =\widetilde{\psi}_{\td, k} (\pk, \rcvk), \bar{t}= \tndk$. Then, we have 
\begin{align}
(1/\tnd)\widetilde{R}_{\tnd,k}(\bp, \rcvk) 
&\geq \widetilde{a}_{\tnd, k}^{(\kappa)} + \widetilde{b}_{\tnd, k}^{(\kappa)} \Big{(}2-\frac{\pk_k}{p_k} - \frac{\widetilde{\psi}_{\td, k} (\bp, \rcvk)}{\widetilde{\psi}_{\td, k} (\pk, \rcvk)}\Big{)} - \widetilde{c}_{\tnd, k}^{(\kappa)} \tnd \triangleq \widetilde{R}_{\tnd, k}^{(\kappa)} (\bp, \rcvk),  \notag
\end{align}
where
\begin{eqnarray}
\ds 0 < &\widetilde{a}_{\tnd,k}^{(\kappa)}=& \frac{2}{\tndk} \ln(1+\frac{\pk_k |h_{k}|^2}{\widetilde{\psi}_{\td, k} (\pk, \rcvk)}), \nonumber\\
\ds 0 < &\widetilde{b}_{\tnd,k}^{(\kappa)}=&  \frac{\pk_k |h_{k}|^2 / \widetilde{\psi}_{\td, k} (\pk, \rcvk)} {\tndk(1+\pk |h_{k}|^2 / \widetilde{\psi}_{\td, k} (\pk, \rcvk))}, \nonumber \\
\ds 0 < &\widetilde{c}_{\tnd,k}^{(\kappa)}=& \frac{\ln(1+\pk_k |h_{k}|^2 / \widetilde{\psi}_{\td, k} (\pk, \rcvk))}{(\tndk)^2}. \nonumber 
\end{eqnarray}

Now, we can reformulate sub-problem 1 as follows:
\begin{subequations} \label{opo3-s1}
	\begin{align}
	&\ds\max_{\bw, \bv, \bp, \btau=(\tni,\tne,\tnd)\in \mathbb{R}_{+}^3} f_{sub-1}^{(\kappa)}(\bw, \bv, \bp, \btau, \rcvk) \triangleq  \min_{d_i \in \clU_I} \widetilde{R}_{\tni, d_i}^{(\kappa)} (\bw, \rcvk) + \eta \Omega(\rcvk) \label{opo3a-s1} \\
	\text{s.t. } \;  &\eqref{opo2d}-\eqref{opo2g}, \nonumber \\
	& \widetilde{E}_{\tne, e_j}^{(\kappa)} (\bv, \rcvk)  \geq e_{min} \tne / \rho, \forall e_j \in \clU_E , \label{opo3b-s1} \\
	& \widetilde{R}_{\tnd, k}^{(\kappa)} (\bp, \rcvk) \geq R_{k,min}, \forall k \in \clK. \label{opo3c-s1} 
	\end{align}
\end{subequations} 
It can be observed from~\eqref{opo3-s1} that $f_{sub-1}^{(\kappa)}(\bw, \bv, \bp, \btau, \rcvk)$ is concave. In particular, the first term in \eqref{op3a-s1} is concave, i.e., the minimum of concave functions \cite{Tuybook}, and the second term is already concave. Therefore, instead of \eqref{op3-s1}, we solve the convex optimization problem given in \eqref{opo3-s1}, and its solution is also used as $(\wkk, \vkk, \pkk, \taukk)$ in the next iteration. The computational complexity of the algorithm to solve the problem in \eqref{opo3-s1} is determined by \eqref{com_complexity}, where $\alpha=2M+K+3$ and $\beta=U_I+2(U_E+K+1)$.

\subsection{Sub-problem 2}
Sub-problem 2 is defined as follows
\begin{subequations} \label{opo2-s2}
	\begin{align}
	&\ds\max_{\brcv} f_{sub-2}(\wk, \vk, \pk, \tauk, \brcv)  \triangleq  \min_{d_i \in \clU_I} (1/\tnik) \widetilde{R}_{\tni, d_i} (\wk, \brcv) + \eta \Omega(\brcv) \label{opo2a-s2} \\
	\text{s.t. } \;  &\eqref{umc1}, \nonumber \\
	& (1/\tnek) \rho \widetilde{E}_{\tne, e_j} (\vk,\brcv) \geq e_{min}, \forall e_j \in \clU_E , \label{opo2b-s2} \\
	& (1/\tndk)\widetilde{R}_{\tnd,k}(\pk, \brcv) \geq R_{k,min}, \forall k \in \clK. \label{opo2c-s2} 
	\end{align}
\end{subequations} 

Similar to Section~\ref{sub-problem1}, to solve~\eqref{opo2-s2}, we convert the objective function and the constraints in~\eqref{opo2b-s2} and \eqref{opo2c-s2} as follows. We apply the inequalities in \eqref{ie1} in the Appendix to the term of $\widetilde{R}_{\tni, d_i} (\wk, \brcv)$ in the objective function with $x = |\bh_{(B,d_i)}(\brcv) \wk_{d_i}|^2, y = \widetilde{\psi}_{\ti, d_i} (\wk, \brcv)$, $\bar{x} = |\bh_{(B,d_i)}(\rcvk) \wk_{d_i}|^2, \bar{y} = \widetilde{\psi}_{\ti, d_i} (\wk, \rcvk)$. Then, we have 
\begin{align}
&\widetilde{R}_{\tni, d_i} (\wk, \brcv) \nonumber \\
&\geq \widetilde{a}_{\tni, d_i}^{(\kappa)} + \widetilde{b}_{\tni, d_i}^{(\kappa)} \Big{(}2-\frac{|\bh_{(B,d_i)}(\rcvk) \wk_{d_i}|^2}{2 \Re \{(\bh_{(B,d_i)}(\brcv) \wk_{d_i})(\bh_{(B,d_i)}(\rcvk) \wk_{d_i})^{*}\} - |\bh_{(B,d_i)}(\rcvk) \wk_{d_i}|^2} - \frac{\widetilde{\psi}_{\ti, d_i} (\wk, \brcv)}{\widetilde{\psi}_{\ti, d_i} (\wk, \rcvk)}\Big{)} \nonumber \\
& \triangleq \widetilde{R}_{\tni, d_i}^{(\kappa)} (\wk, \brcv)
\end{align}
over the trust region $2 \Re \{(\bh_{(B,d_i)}(\brcv) \wk_{d_i})(\bh_{(B,d_i)}(\rcvk) \wk_{d_i})^{*}\} - |\bh_{(B,d_i)}(\rcvk) \wk_{d_i}|^2 \geq 0$, where
\begin{eqnarray}
\ds 0 < &\widetilde{a}_{\tni, d_i}^{(\kappa)}=& \ln\big{(}1+\frac{|\bh_{(B,d_i)}(\rcvk) \wk_{d_i}|^2}{\widetilde{\psi}_{\ti, d_i} (\wk, \rcvk)}\big{)}, \nonumber\\
\ds 0 < &\widetilde{b}_{\tni, d_i}^{(\kappa)}=&  \frac{|\bh_{(B,d_i)}(\rcvk) \wk_{d_i}|^2 / \widetilde{\psi}_{\ti, d_i} (\wk, \rcvk)} {(1+|\bh_{(B,d_i)}(\rcvk) \wk_{d_i}|^2 / \widetilde{\psi}_{\ti, d_i} (\wk, \rcvk))}. \nonumber 
\end{eqnarray}
Moreover, according to \eqref{op3-ob1}, we can express $\Omega(\brcv)$ as 
\begin{align}
\Omega(\brcv) &\geq  \frac{1}{N} - \frac{1}{\sum_{n=1}^{N}  (2 \Re\{(\rcvk_n)^{*} \theta_n\}  - |\rcvk_n|^2)}\triangleq \Omega^{(\kappa)}(\brcv). \nonumber
\end{align}
For the constraint in \eqref{opo2b-s2}, it can be replaced by
\begin{align}
\widetilde{E}_{\tne, e_j} (\vk,\brcv)
&\geq \ds\sum_{l=1}^{U_E} \left[2 \Re \{(\bh_{(B,e_j)}(\brcv) \vk_{e_l})(\bh_{(B,e_j)}(\rcvk) \vk_{e_l})^{*}\} - |\bh_{(B,e_j)}(\rcvk) \vk_{e_l}|^2 \right] \nonumber \\
&\triangleq \widetilde{E}_{\tne, e_j}^{(\kappa)} (\vk, \brcv)  \geq e_{min} \tnek / \rho.
\end{align}
For the constraint in \eqref{opo2c-s2}, we also apply the inequalities in \eqref{ie1} in the Appendix to $\widetilde{R}_{\tnd,k}(\pk, \brcv)$ with $x = \pk |h_{k}(\brcv)|^2, y = \widetilde{\psi}_{\td, k} (\pk, \brcv)$ and $\bar{x} = \pk |h_{k}(\rcvk)|^2, \bar{y} = \widetilde{\psi}_{\td, k} (\pk, \rcvk)$. We have
\begin{align}
\widetilde{R}_{\tnd,k}(\pk, \brcv) 
&\geq \widetilde{a}_{\tnd, k}^{(\kappa)} + \widetilde{b}_{\tnd, k}^{(\kappa)} \Big{(}2-\frac{|h_{k}(\rcvk)|^2}{2 \Re \{(h_{k}(\brcv))(h_{k}(\rcvk))^{*}\} - |h_{k}(\rcvk)|^2}-\frac{\widetilde{\psi}_{\td, k} (\pk, \brcv)}{\widetilde{\psi}_{\td, k} (\pk, \rcvk)}\Big{)} \nonumber \\
& \triangleq \widetilde{R}_{\tnd, k}^{(\kappa)} (\pk, \brcv) 
\end{align}
over the trust region $2 \Re \{(h_{k}(\brcv))(h_{k}(\rcvk))^{*}\} - |h_{k}(\rcvk)|^2 \geq 0$, where
\begin{eqnarray}
\ds 0 < &\widetilde{a}_{\tnd,k}^{(\kappa)}=& \ln(1+\frac{\pk |h_{k}|^2}{\widetilde{\psi}_{\td, k} (\pk, \rcvk)}), \nonumber\\
\ds 0 < &\widetilde{b}_{\tnd,k}^{(\kappa)}=&  \frac{\pk |h_{k}(\rcvk)|^2 / \widetilde{\psi}_{\td, k} (\pk, \rcvk)} {1+\pk |h_{k}(\rcvk)|^2 / \widetilde{\psi}_{\td, k} (\pk, \rcvk)}. \nonumber
\end{eqnarray}

From the above derivations, we can reformulate sub-problem 2 as follows
\begin{subequations} \label{opo3-s2}
	\begin{align}
	&\ds\max_{\brcv} f_{sub-2}^{(\kappa)}(\wk, \vk, \pk, \tauk, \brcv) \triangleq  \min_{d_i \in \clU_I} (1/\tndk) \widetilde{R}_{\tnd, d_i}^{(\kappa)} (\wk, \brcv)  + \eta \Omega^{(\kappa)}(\brcv) \label{opo3a-s2} \\
	\text{s.t. } \;  &\eqref{umc1}, \nonumber \\
	& \widetilde{E}_{\tne, e_j}^{(\kappa)} (\vk, \brcv)  \geq e_{min} \tnek / \rho, \forall e_j \in \clU_E , \label{opo3b-s2} \\
	& (1/\tndk)\widetilde{R}_{\tnd,k}^{(\kappa)}(\pk, \brcv) \geq R_{k,min}, \forall k \in \clK. \label{opo3c-s2} 
	\end{align}
\end{subequations} 

The problem in \eqref{opo3-s2} is now convex, and the computational complexity of the algorithm to solve this problem is determined by \eqref{com_complexity}, where $\alpha=N$ and $\beta=N+U_E+K$. The overall algorithm to solve the optimization problem in \eqref{opo2} is presented in Algorithm~\ref{Alg_2}. Similar to Algorithm~\ref{Alg_1}, it is important to generate a feasible point as well as to select the penalty parameter $\eta$ for Algorithm~\ref{Alg_2}, which is omitted here due to the paper length constraint.

\begin{algorithm}
\footnotesize
         \caption{\footnotesize OTA Algorithm for solving the problem in \eqref{opo2}}\label{Alg_2}
        \begin{algorithmic}[1]
 \State Initialize: Find a feasible point $(\wk, \vk, \pk, \tau^{(0)}=(\tni^{(0)},\tne^{(0)}, \tnd^{(0)}), \theta^{(0)})$ for the problem in \eqref{opo2}. Set $\kappa = 0$;
         	\State Compute $\eta$ according to \eqref{eta_1} from the obtained feasible set;
           \Repeat 
		\State Solve the convex problem in \eqref{opo3-s1} for $\brcv = \theta^{(\kappa)}$ to generate $(\wkk, \vkk, \pkk, \taukk)$;
		\State Solve the convex problem in \eqref{opo3-s2} for $(\bw, \bv, \bp, \btau)=(\wkk, \vkk, \pkk, \taukk)$ to generate $\theta^{(\kappa+1)}$;
		\State $\kappa \gets \kappa+1$;
	\Until Convergence.
	\State Output $(\wk, \vk, \pk, \tauk=(\tni^{(\kappa)}, \tne^{(\kappa)}, \tnd^{(\kappa)}), \rcvk)$ 
       \end{algorithmic}
    \end{algorithm}

\section{Performance Evaluation}
\label{perform_eval}
In this section, we present and discuss simulation results obtained by the proposed algorithms. For an ease of presenting the results, we consider a scenario that consists of one BS, one IRS and two IUs, two EUs and three D2D pairs. Nevertheless, the algorithms can scale efficiently to bigger network sizes. The coordinates of the BS and IRS are at $(40, 0, 25)$ and $(0, 60, 40)$, respectively. Meanwhile, the IUs, EUs and D2D pairs are randomly located in an area of $120\rm{m} \times 120\rm{m}$. We consider the large-scale fading between the BS and IRS with an coefficient $\gamma_{BS-IRS}$ (in dB) determined as $\gamma_{BS-IRS}=G_{BS}+G_{IRS}-35.9-22\log_{10}(d_{BS-IRS})$, where $G_{BS}=5$ dBi is the antenna gain of the BS, $G_{IRS}=5$ dBi is the gain of the IRS elements, respectively, and $d_{BS-IRS}$ is the distance between the BS and IRS. Similar to~\cite{kammoun2020asymptotic}, we assume a full-rank BS-to-IRS LoS channel matrix that is defined as $[\bG]_{n,m}=e^{j\pi\big((n-1)\sin\bar{\theta}_{LoS}(n) \sin\bar{\phi}_{LoS}(n)+(m-1)\sin\theta_{LoS}(n) \sin \phi_{LoS}(n)\big)}$, where $\theta_{LoS}(n)$ and $\phi_{LoS}(n)$ are uniformly distributed as $\theta_{LoS}(n) \sim \mathcal{U}(0,\pi)$ and $\phi_{LoS}(n) \sim \mathcal{U}(0,2\pi)$, respectively, and $\bar{\theta}_{LoS}(n)=\pi-\theta_{LoS}(n)$ and $\bar{\phi}_{LoS}(n)=\pi+\phi_{LoS}(n)$. The channel from the BS to each user at a distance of $d$ meters is determined by $\sqrt{10^{-\alpha_{PL}/10}}\tilde{h}$, where  $\tilde{h}$ is the channel gain from the BS to the user, and $\alpha_{PL}$ is the path-loss of the channel. In particular, for the EUs, $\tilde{h}$ is the Rician fading channel gain associated with a Rician factor of $10$ dB, and for other users, i.e., the IUs and D2D users, $\tilde{h}$ is the normalized Rayleigh fading channel gain. The path-loss $\alpha_{PL}$ (in dB) is determined as $\alpha_{PL}=30+10\gamma\log_{10}(d)$, where $\gamma$ is the path-loss exponent. In particular, $\gamma$ is set to $3$ for the Rician channels and is set to $2$ for the Rayleigh channels. Other simulation parameters are provided in Table~\ref{table:parameters} that are similar to those in \cite{IEEE_IRS_HDTuan}.
\begin{table}[h!]
\caption{\small Simulation parameters.}
\label{table:parameters_CRN}
\footnotesize 	
\centering
\begin{tabular}{lc|lc|lc|lc}
\hline\hline
{\em Parameters} & {\em Value} & {\em Parameters} & {\em Value}  & {\em Parameters} & {\em Value}&   {\em Parameters} & {\em Value}\\ [1ex]
\hline
 $M$    & $6$           &    $N$ &   $10$   &    $\rho$ &   $0.5$   &    $e_{min}$   &  $0$ dBm        \\ 
\hline
 $U_I$    & $2$          &       $U_E$& $2$ &    $K $  & $3$     &  $R_{k,min}$  &     $0.4$ bps/Hz       \\ 
\hline
 $B$ & $10$ MHz          &       $P_{B,max}$ & $20$ dBm &     $P_{k,max}$ &   $20$ dBm&   $\noise_{d_i},\noise_{k}$   & -174 dBm/Hz   \\ 
\hline
\end{tabular}
\label{table:parameters}
\end{table}

For the performance comparison, we consider the following algorithms:
\begin{itemize}
\item \textbf{N-OTA:} This is the alternating descent algorithm as described in Section III (shown in Algorithm \ref{Alg_1}), which is proposed to solve the optimization problem given in \eqref{op2}. 
\item \textbf{N-OTA with random $\theta$:} This algorithm is similar to the N-OTA algorithm in which the phase shifts of the IRS are random. 
\item \textbf{OTA:} This is the alternating descent algorithm as described in Section IV (shown in Algorithm \ref{Alg_2}), which is proposed to solve the optimization problem given in \eqref{opo2}. 
\item \textbf{OTA with random $\theta$:} This is similar to the OTA algorithm in which the phase shifts of the IRS are random.
\end{itemize}

\begin{figure*}[h]
  \centering
  \subcaptionbox{}[.45\linewidth][c]{%
    \includegraphics[width=0.85\linewidth]{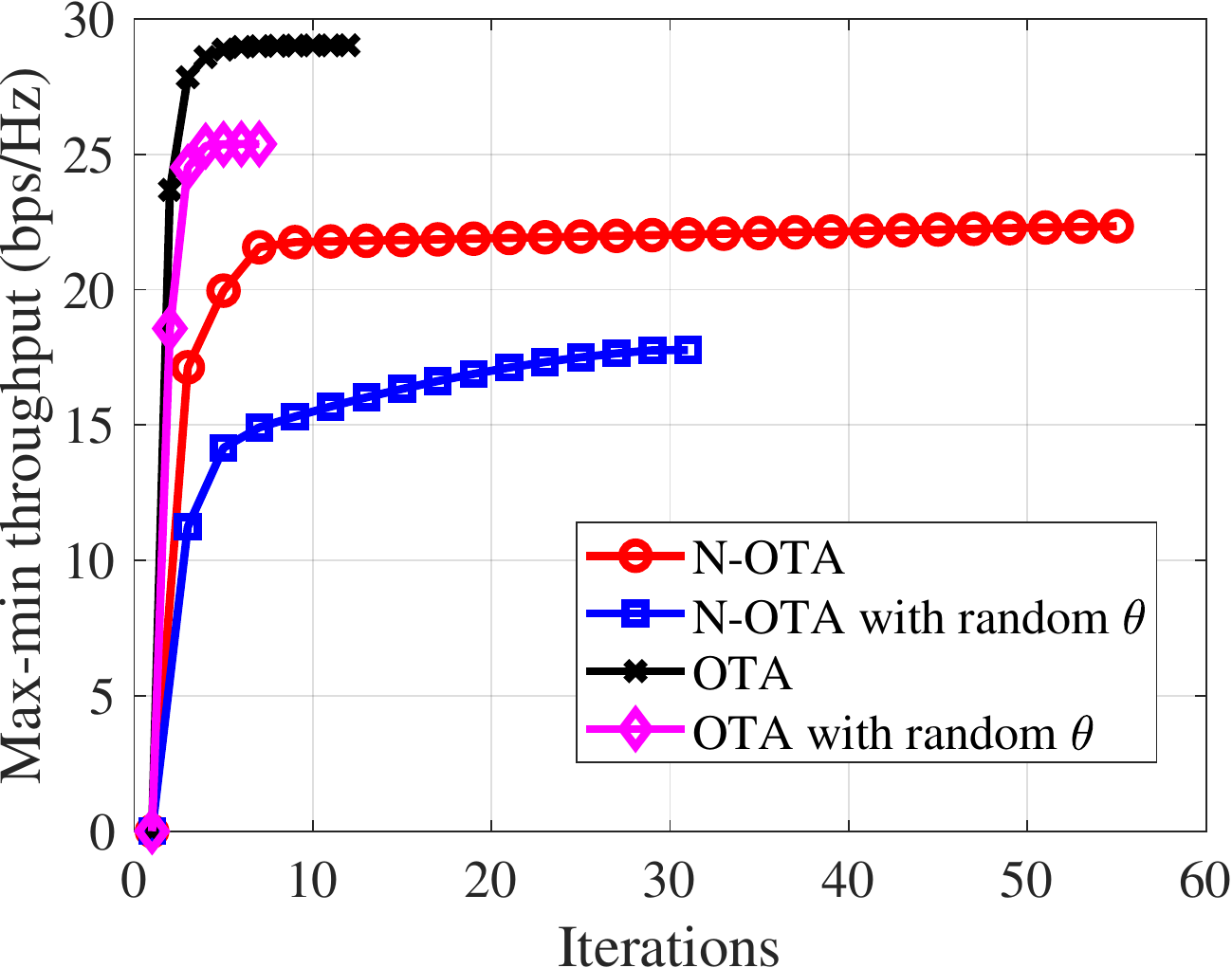}}\quad
  \subcaptionbox{}[.45\linewidth][c]{%
    \includegraphics[width=0.85\linewidth]{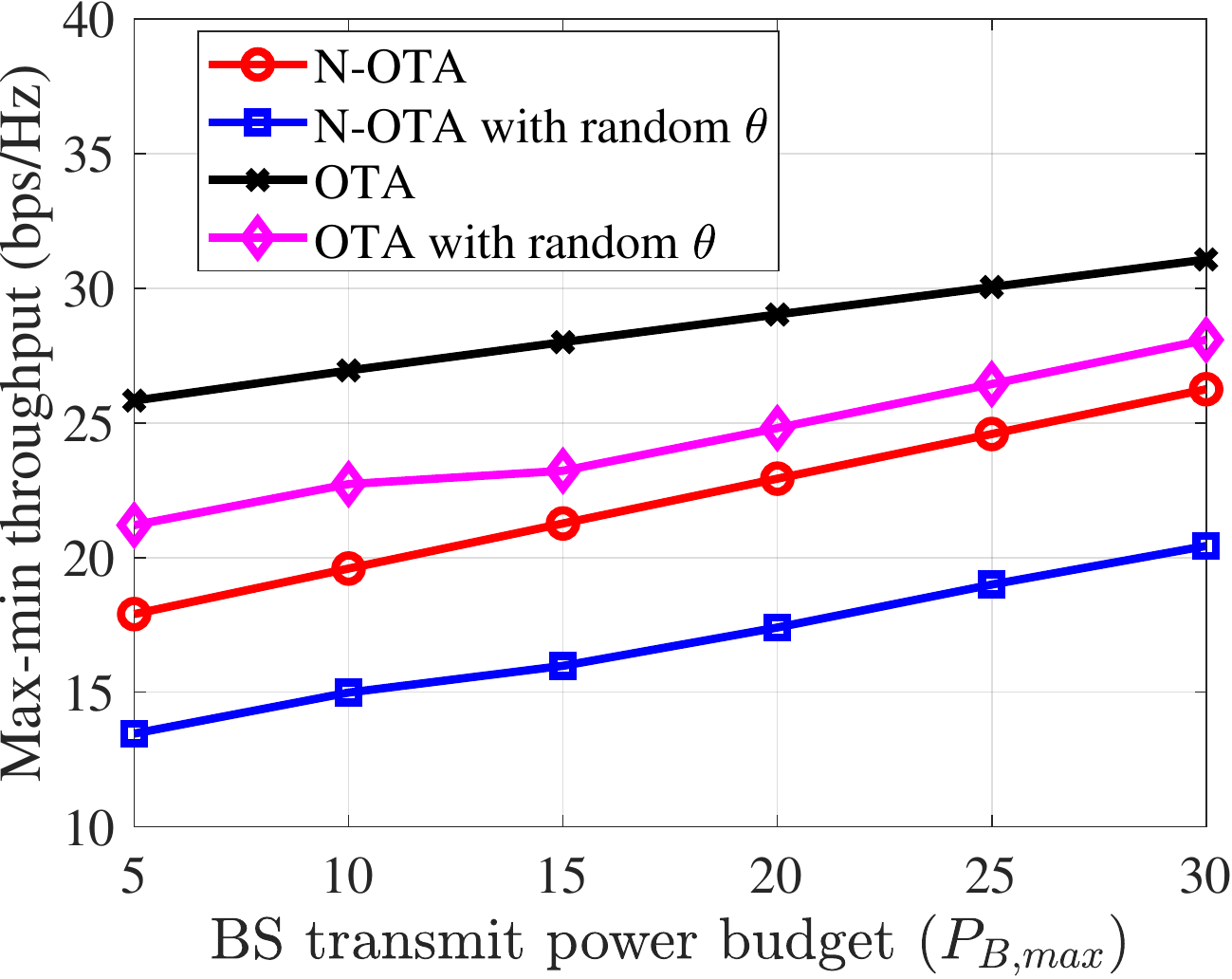}}
\caption{a) The convergence of the algorithms and b) the max-min throughput versus BS's power budget $P_{B,max}$.}
	\label{iter_max_power}
\end{figure*}

First, we discuss the convergence of the algorithms and the max-min throughput obtained by the algorithms. As shown in Fig.~\ref{iter_max_power}(a), all the algorithms are able to  rapidly converge to their stable values. As seen, the OTA algorithm converges faster than the N-OTA algorithm. Moreover, max-min throughput obtained by the IUs with the OTA algorithm is higher than that with the N-OTA algorithm. This is due to the fact that there exists interference caused by the D2D communications to the IUs in the N-OTA scenario. This reduces the SINR at the IUs and decreases their throughput. It can also be seen from Fig.~\ref{iter_max_power}(a) that the max-min throughputs obtained by the OTA and N-OTA are much higher than those obtained by the OTA and N-OTA with random $\theta$, respectively. This demonstrates the effectiveness of our proposed algorithms.


Next, we discuss the impact of the maximum transmit power, i.e., $P_{B,max}$, of the BS on the max-min throughput achieved by the IUs. As shown in Fig.~\ref{iter_max_power}(b), as  $P_{B,max}$ increases, the max-min throughput of IUs increases. This is because of that the data throughput of each IU is proportional to $P_{B,max}$. Moreover, it can be seen from the figure that over the values of $P_{B,max}$, the max-min throughputs obtained by the proposed algorithms with phase shift optimization are always higher than those obtained by the baseline algorithms with random $\theta$. For example, the N-OTA algorithm improves the max-min throughput up to $22$\% compared with the N-OTA algorithm with random $\theta$.


\begin{figure*}[h]
  \centering
  \subcaptionbox{}[.45\linewidth][c]{%
    \includegraphics[width=0.85\linewidth]{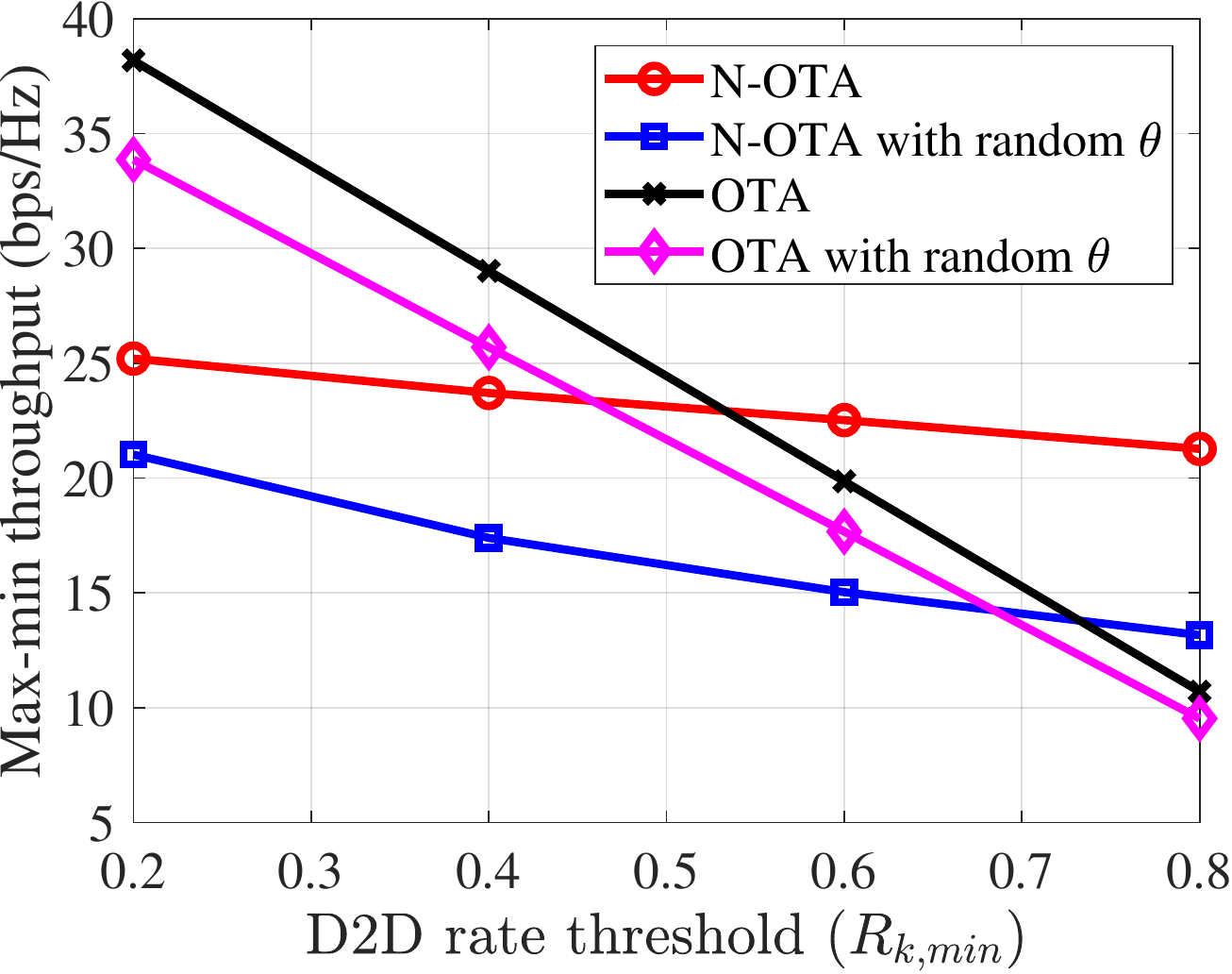}}\quad
  \subcaptionbox{}[.45\linewidth][c]{%
    \includegraphics[width=0.85\linewidth]{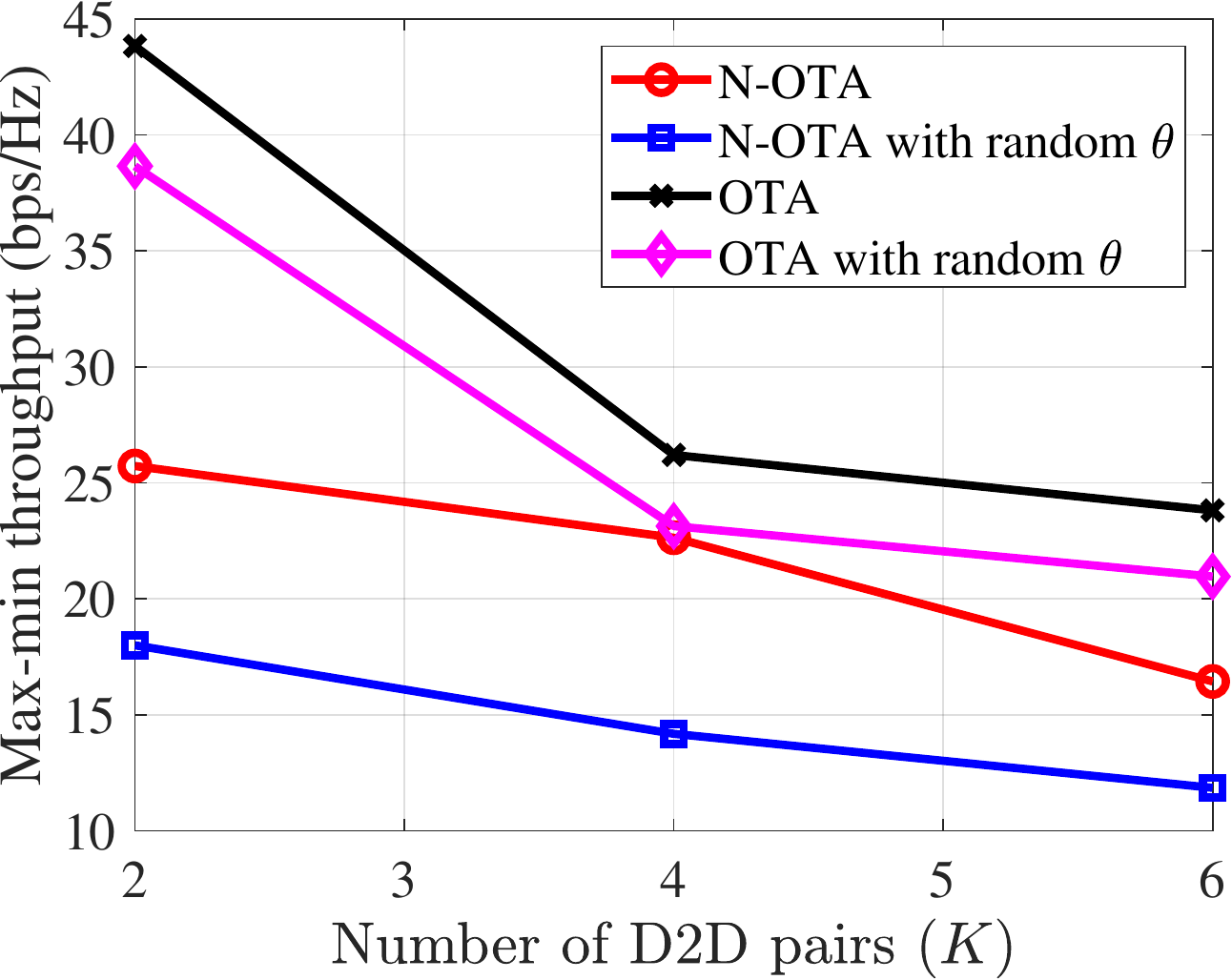}}
\caption{a) Achievable max-min throughput versus (a) the D2D rate threshold $R_{k,min}$ and b) the number of D2D pairs $K$}
	\label{change_rmin}
\end{figure*}

Since the integrated data and energy network coexists with the D2D communications, it is important to show the impact of the D2D rate threshold, i.e., $R_{k,min}$, on the max-min throughput achieved by the IUs. As shown in Fig.~\ref{change_rmin}(a), as $R_{k,min}$ increases, the throughputs obtained by all the algorithms decrease. This result can be explained as follows. With the N-OTA algorithm, as $R_{k,min}$ increases, the D2D transmitters need to transmit their signals with higher power. This increases the interference from D2D communications to the IUs, which reduces the SINR at IUs and decreases their throughputs. With the OTA algorithm, as $R_{k,min}$ increases, more time is allocated to the D2D communications, which reduces the time assigned to the IUs and decreases their throughputs. It is interesting that as $R_{k,min}$ increases, the throughput obtained by the OTA algorithm decreases faster than that obtained by the N-OTA algorithm. The reason may be that the phase shift optimization of the N-OTA algorithms helps to reduce the interference caused by the D2D communications to the IUs. Therefore, as $R_{k,min}$ increases, the increase of interference caused by the D2D communications may not be significant, and the throughput obtained by the N-OTA algorithm slowly decreases. Thus, as $R_{k,min}$ is high, i.e., $\geq 0.7$ bps/Hz, the throughput obtained by the N-OTA algorithm is higher than that obtained by the OTA algorithm. These results further imply that to achieve the high throughput for the IUs, the OTA scenario can be considered when $R_{k,min}$ is low, and the N-OTA scenario is considered when $R_{k,min}$ is high.


Now, we discuss the impact of number of D2D pairs $K$ on the max-min throughput of IUs, and the results are shown in Fig.~\ref{change_rmin}(b). As seen, as $K$ increases, the throughputs obtained by both the N-OTA and OTA algorithms decrease. In particular for the N-OTA algorithm, the reason is that as $K$ increases, the interference caused by the D2D communications to the IUs increases. For the OTA algorithm, as $K$ increases, the SINR at each D2D receiver reduces due to the increase of interference from the D2D transmitters. To satisfy the fixed D2D rate threshold, more time is required (and allocated) to the D2D communications. As a result, this reduces the time allocated to the IUs and their throughputs. Note that over the values of $K$, the max-min throughput with the phase shift optimization is always higher than that with the random $\theta$.


\begin{figure*}[h]
  \centering
  \subcaptionbox{}[.45\linewidth][c]{%
    \includegraphics[width=0.85\linewidth]{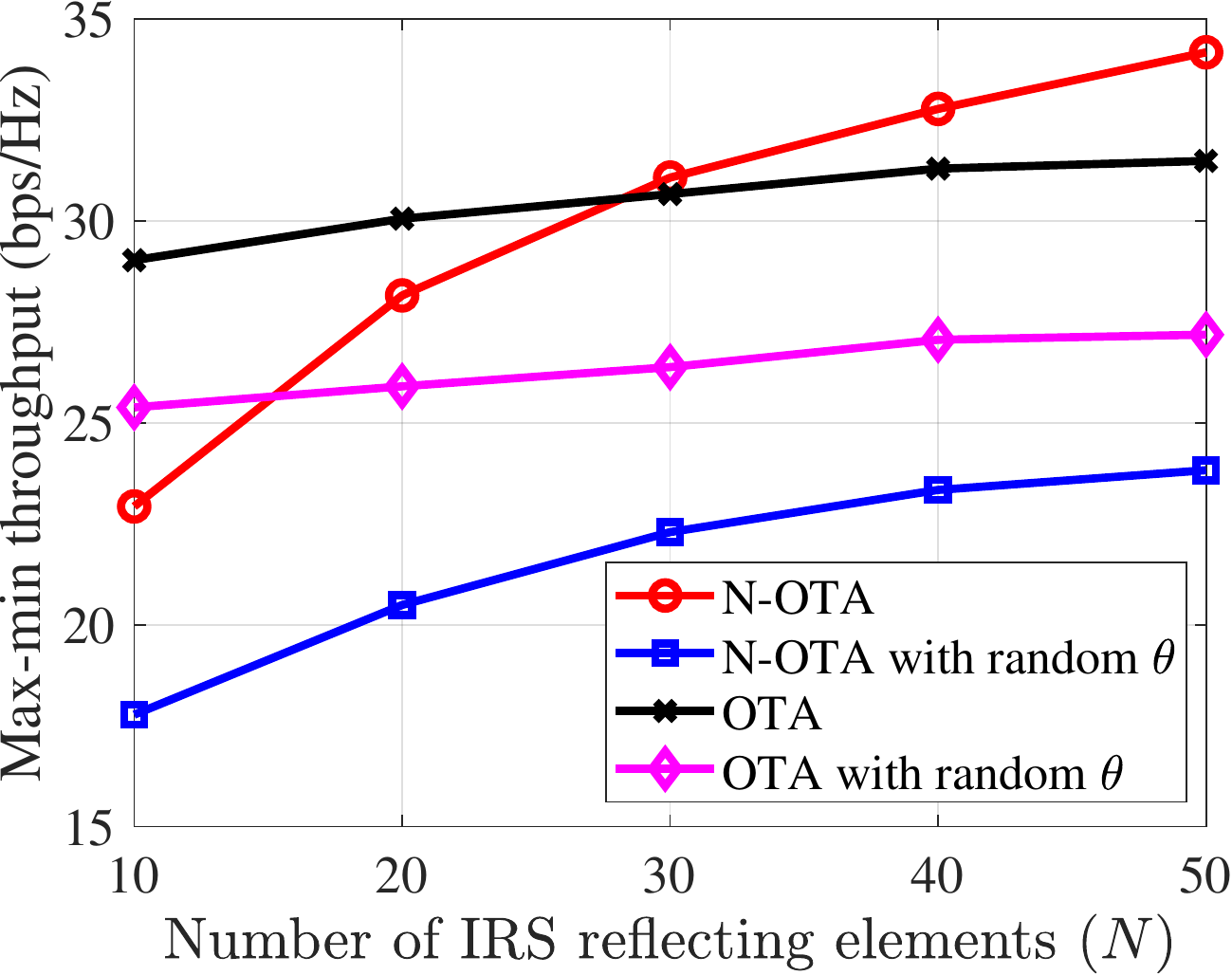}}\quad
  \subcaptionbox{}[.45\linewidth][c]{%
    \includegraphics[width=0.85\linewidth]{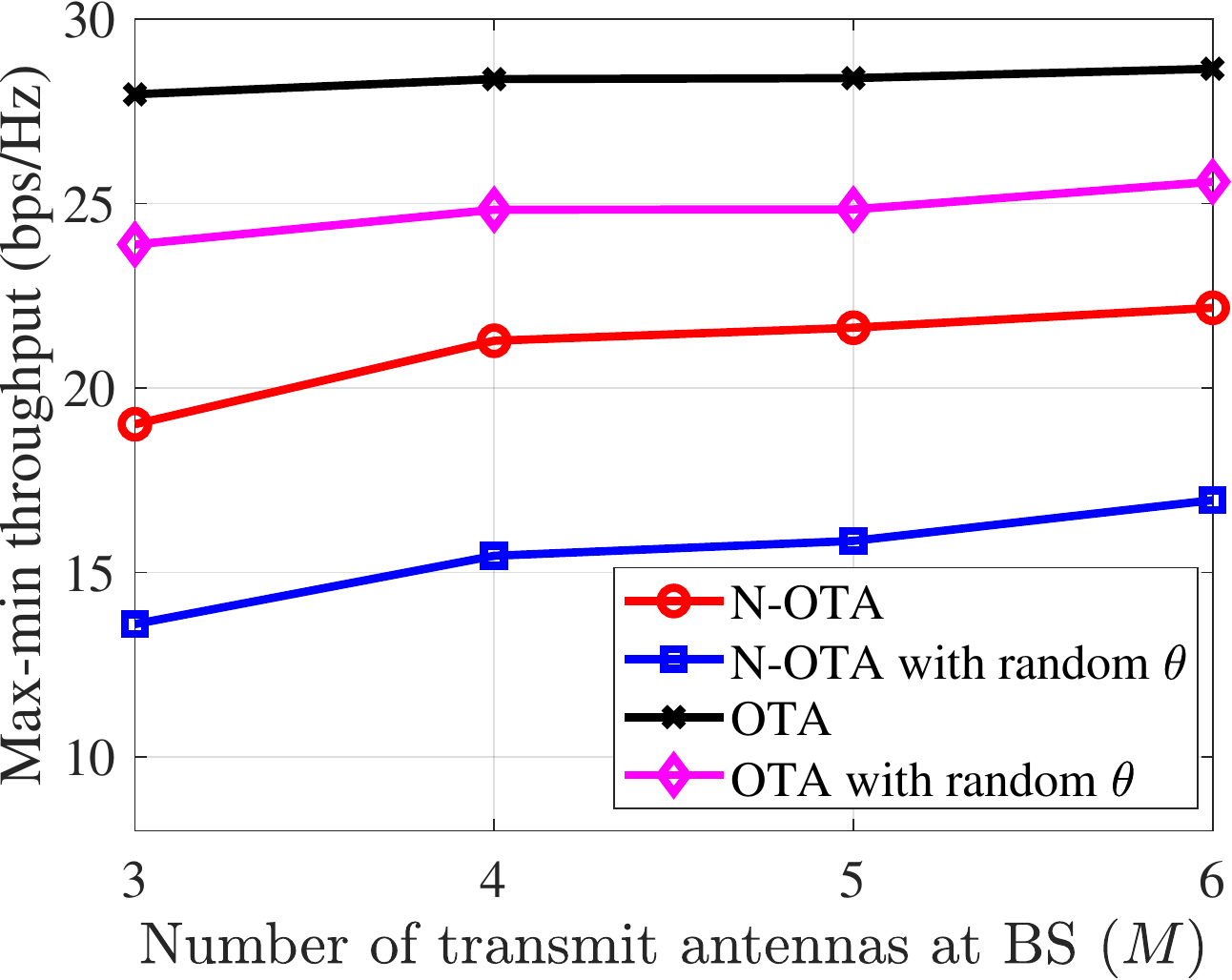}}
\caption{Achievable max-min throughput versus a) the number of IRS elements and b) the number of antennas at the BS.}
	\label{change_irs_anten}
\end{figure*}

It is worth noting that the size of the IRS, i.e., $N$, can also affect the max-min throughput achieved by the IUs. As shown in Fig.~\ref{change_irs_anten}(a), as $N$ increases, the max-min throughputs achieved by both the N-OTA and OTA algorithms increase. The reason is that the SINR at the IUs increases as $N$ increases. However, the the increasing rate of the N-OTA algorithms is faster that of the OTA algorithm. This can explained based on the definition of the throughput of the IUs (in Sections III and IV). Indeed, in both the N-OTA and OTA scenarios, the throughput of each IU is in the form of $\ti R_{\ti, d_i}$, where $\ti$ is the time allocated to IU $i$ and $R_{\ti, d_i}$ is the data rate obtained by the IU. In the N-OTA scenario, there is no time allocated to the D2D communications, and thus $\ti$ is larger than that in the OTA scenario. Therefore, as $N$ increases, $R_{\ti, d_i}$ obtained by the N-OTA algorithm increases faster than that obtained by the OTA algorithm. The throughput of the IUs also increases as the number of antennas of the BS increases as shown in Fig.~\ref{change_irs_anten}(b). This is obvious since the SINR at the IUs increases. 



\section{Conclusions}
\label{sec:conc}
In this paper, we have considered the IRS-aided DED2D system in which the integrated data and energy network and D2D communication coexist with the assistance of IRS. We have formulated the max-min throughput optimization problem with the aim of maximizing the minimum throughput of the IUs, subject to the harvested energy requirement of the EUs and the data rate threshold of the D2D communications. The max-min throughput optimization problem is computationally intractable, and we have proposed the alternating descent algorithm to solve it. We have further considered the max-min throughput optimization problem in the OTA scenario in which there is the time allocation to the D2D communications. We have provided the simulation results to evaluate and compare the effectiveness of the proposed algorithms. 

\section*{Appendix~A: Fundamental inequalities} \label{Appendix:A}
\renewcommand{\theequation}{A.\arabic{equation}}
\setcounter{equation}{0}
Function $f(x) = \ln(1+1/x)$ is convex on the domain $x>0$
\begin{align} \label{ie0}
\ln(1+\frac{1}{x}) \geq \ln(1+\frac{1}{\bar{x}}) + \frac{1}{1+\bar{x}}\left(1 - \frac{x}{\bar{x}}\right).
\end{align}

The convex function $f(x,y) = \ln(1+\frac{1}{x y})$ on $\mathbb{R}_{+}^2$
\begin{align} \label{ie0-1}
\ln(1+\frac{1}{x y}) &\geq  f(\bar{x},\bar{y}) + \la \nabla f(\bar{x},\bar{y}), (x,y) - (\bar{x},\bar{y}) \ra \nonumber \\
&\geq \ln(1+1/\bar{x} \bar{y}) + \frac{1/\bar{x} \bar{y}} {\bar{t}(1+1/\bar{x} \bar{y})} (2-\frac{x}{\bar{x}}-\frac{y}{\bar{y}}).
\end{align}
Substituting $x\rightarrow1/x$ and $\bar{x}\rightarrow1/\bar{x}$
\begin{align} \label{ie1}
\ln(1+\frac{x}{y}) \geq \ln(1+\bar{x} / \bar{y}) + \frac{\bar{x} / \bar{y}} {(1+\bar{x} / \bar{y})} (2-\frac{\bar{x}}{x}-\frac{y}{\bar{y}}).
\end{align}

From \cite{sheng2018power}, the convex function $f(x,y,t) = \frac{1}{t}\ln(1+\frac{1}{x y})$ on $\mathbb{R}_{+}^3$. Therefore 
\begin{align} \label{ie0-2}
\frac{1}{t}\ln(1+\frac{1}{x y}) &\geq  f(\bar{x},\bar{y},\bar{t}) + \la \nabla f(\bar{x},\bar{y},\bar{t}), (x,y,t) - (\bar{x},\bar{y},\bar{t}) \ra \nonumber \\
&\geq \frac{2}{\bar{t}} \ln(1+1/\bar{x} \bar{y}) + \frac{1/\bar{x} \bar{y}} {\bar{t}(1+1/\bar{x} \bar{y})} (2-\frac{x}{\bar{x}}-\frac{y}{\bar{y}}) - \frac{\ln(1+1/\bar{x} \bar{y})}{\bar{t}^2} t.
\end{align}
Substituting $x\rightarrow1/x$ and $\bar{x}\rightarrow1/\bar{x}$
\begin{align} \label{ie2}
&\frac{1}{t}\ln(1+\frac{x}{y}) \geq \frac{2}{\bar{t}} \ln(1+\bar{x} / \bar{y}) + \frac{\bar{x} / \bar{y}} {\bar{t}(1+\bar{x} / \bar{y})} (2-\frac{\bar{x}}{x}-\frac{y}{\bar{y}}) - \frac{\ln(1+\bar{x} / \bar{y})}{\bar{t}^2} t.
\end{align}

\bibliographystyle{IEEEtran}
\bibliography{IEEEabrv,Paper_Bib}

\end{document}